\documentclass[12pt]{article}

\usepackage{amssymb}
\usepackage[mathscr]{euscript}
\usepackage{graphicx}

\newtheorem{prop}{Proposition}[section]

\newtheorem{defi}{Definition}[section]
\newtheorem{lemm}{Lemma}[section]
\newtheorem{theo}{Theorem}[section]
\newtheorem{coro}{Corollary}[section]

\newcommand{\bbox}{\normalsize {}%
        \nolinebreak \hfill $\blacksquare$ \medbreak \par}

\newcommand{\Ni}{\hbox{ {\vrule height .22cm}{\leaders\hrule\hskip.2cm} }}
\newcommand{\iN}{\hbox{ {\leaders\hrule\hskip.2cm}{\vrule height .22cm} }}

\newcommand{\R}[1][]{\ensuremath{{\mathbb{R}^{#1}} }}

\def\<{\langle} \def\>{\rangle}

\title{The notion of observable in the covariant Hamiltonian
formalism for the calculus of variations with several variables}
\author{
Fr\'ed\'eric H\'ELEIN\footnote{helein@math.jussieu.fr} \\
Institut de Mathématiques de Jussieu, UMR 7586,\\
Université Denis Diderot--Paris 7, Site de Chevaleret,\\
16 rue Clisson 75013 Paris
(France)
\\ \\
Joseph KOUNEIHER\footnote{kouneiher@paris7.jussieu.fr}\\
LUTH, CNRS UMR 8102\\ Observatoire de Paris - section Meudon \\5 Place Jules Janssen \\92195 Meudon Cedex \\ Universit\'e Paris 7\\
}

\begin{document}
\maketitle
\begin{center}
{\bf Abstract}
\end{center}
This papers is concerned with multisymplectic formalisms which are
the frameworks for Hamiltonian theories for fields theory. Our
main purpose is to study the observable $(n-1)$-forms which allows
one to construct observable functionals on the set of solutions of
the Hamilton equations by integration. We develop here two
different points of view: generalizing the law $\{p,q\} = 1$ or
the law $dF/dt = \{H,F\}$. This leads to two possible
definitions; we explore the relationships and the differences
between these two concepts. We show that --- in contrast with the
de Donder--Weyl theory --- the two definitions coincides in the Lepage--Dedecker theory.

\section{Introduction}
Multisymplectic formalisms are the frameworks for finite dimensional formulations of
variational problems with several variables (or field theories for
physicists) analogous to the well-known Hamiltonian theory of
point mechanics. They are based on the following analogues of symplectic forms:
given a differential manifold ${\cal M}$ and $n\in
\Bbb{N}$ a smooth $(n+1)$-form $\Omega$ on ${\cal M}$ is a {\em
multisymplectic} form if and only if $\Omega$ is non degenerate
(i.e.\,$\forall m\in {\cal M}$, $\forall \xi \in T_m{\cal M}$, if
$\xi \iN \Omega_m = 0$, then $\xi = 0$) and closed. We call
$({\cal M},\Omega)$ a {\em multisymplectic} manifold. Then
one can associates to a Lagrangian variational problem a multisymplectic
manifold $({\cal M}, \Omega)$ and a Hamiltonian function ${\cal H}$ on ${\cal M}$
s.t.\,any solution of the variational problem is represented by
a solution of a system of generalized Hamilton equations.
Geometrically this solution is pictured by an $n$-dimensional submanifold
$\Gamma\subset {\cal M}$ s.t.\,$\forall m\in {\cal M}$ there
exists a $n$-multivector $X$ tangent to ${\cal M}$ at $m$
s.t.\,$X\iN \Omega = (-1)^nd{\cal H}$. We then call $\Gamma$ a
{\em Hamiltonian $n$-curve}.\\

\noindent For example a variational problem on maps
$u:\Bbb{R}^n\longrightarrow \Bbb{R}$ can written as the system\footnote{an
obvious difference with  mechanics is that
there is a dissymmetry between the ``position'' variable $u$ and
the ``momentum'' variables $p^\mu$. Since (\ref{0.h}) involves a
divergence of $p^\mu$ one can anticipate that, when formulated in
more geometrical terms, $p^\mu$ will be interpreted as the
components of a $(n-1)$-form, whereas $u$ as a scalar function.}
\begin{equation}\label{0.h}
{\partial u\over \partial x^{\mu}} =
{\partial H\over \partial p^{\mu}}(x,u,p)
\quad \hbox{ and }\quad
\sum_{\mu}{\partial p^{\mu}\over \partial x^{\mu}}
= - {\partial H\over \partial u}(x,u,p).
\end{equation}
The corresponding
multisymplectic form is $\Omega:=d\theta$, where $\theta:=  e \omega + p^{\mu} du\wedge
\omega_{\mu}$ (and $\omega:= dx^1\wedge \cdots  \wedge dx^n$ and
$\omega_{\mu}:= \partial _\mu\iN \omega$). The simplest example
of such a theory was proposed by T. de Donder \cite{deDonder} and H. Weyl \cite{Weyl}. But it
is a particular case of a huge variety of
multisymplectic theories which were discovered by T. Lepage and
can be described using a universal framework built by P.
Dedecker \cite{Dedecker}, \cite{HeleinKouneiher}, \cite{HK1a}.\\

\noindent The present paper, which is a continuation of
\cite{HeleinKouneiher} and \cite{HK1a}, is devoted to the study of
observable functionals defined on the set of all Hamiltonian
$n$-curves $\Gamma$. An important class of such functionals can be
constructed by choosing appropriate $(n-1)$-forms $F$ on the
multisymplectic manifold ${\cal M}$ and a hypersurface $\Sigma$ of
${\cal M}$ which crosses transversally all Hamiltonian $n$-curves
(we shall call {\em slices} such hypersurfaces). Then
$\int_{\Sigma}F:\Gamma \longmapsto \int_{\Sigma\cap \Gamma}F$ is
such a functional. One should however check that such functionals
measure physically relevant quantities. The philosophy adopted
here is inspired from quantum Physics: the formalism should
provide us with rules for predicting the dynamical evolution of an
observable. There are two ways to translate this requirement
mathematically: first the ``infinitesimal evolution'' $dF(X)$ of
$F$ along a $n$-multivector $X$ tangent to a Hamiltonian $n$-curve
should be completely determined by the value of $d{\cal H}$ at the
point --- this leads to the definition of what we call an {\em
observable $(n-1)$-form} (OF), the subject of Section 3;
alternatively, inspired by an analogy with classical particle
mechanics, one can assume that there exists a tangent vector field
$\xi_F$ such that $\xi_F\iN \Omega + dF = 0$ everywhere --- we
call such forms {\em algebraic observable $(n-1)$-forms} (AOF).
This point of view will be investigated in Section 4. We believe
that the notion of AOF was introduced by W. Tulczyjew in 1968
\cite{Tulczyjew-} (see also \cite{Gawedski},
\cite{GoldschmidtSternberg}, \cite{Kijowski1}). To our knowledge
the notion of OF was never considered before; it seems to us
however that it is a more natural definition. It is easy to check
that all AOF are actually OF but the converse is in general not
true (see Section 4), as in particular in the de Donder--Weyl
theory.\\

\noindent It is worth here to insist on the difference of points
of view between choosing OF's or AOF's. The definition of OF is in
fact the right notion if we are motivated by the interplay between
the dynamics and observable functionals. It allows us to define a
{\em pseudobracket} $\{{\cal H},F\}$ between the Hamiltonian
function and an OF $F$ which leads to a generalization of the
famous equation ${dA\over dt} = \{H,A\}$ of the Hamiltonian
mechanics. This is the relation
\begin{equation}\label{0.dF}
dF_{|\Gamma} = \{{\cal H}, F\}\omega_{|\Gamma},
\end{equation}
where $\Gamma$ is a Hamiltonian $n$-curve and $\omega$ is a given
volume $n$-form on space-time (see Proposition 3.1). In contrast
the definition of AOF's is the right notion if we are motivated in
defining an analogue of the Poisson bracket between observable
$(n-1)$-forms. This Poisson bracket, for two AOL $F$ and $G$ is
given by $\{F,G\}:= \xi_F\wedge \xi_G\iN \Omega$, a definition
reminiscent from classical mechanics. This allows us to construct
a Poisson bracket on functionals by the rule $\{\int_\Sigma
F,\int_\Sigma G\}: \Gamma \longmapsto \int_{\Sigma\cap
\Gamma}\{F,G\}$ (see Section 4).\\

\noindent Note that it is possible to generalize the notion of
observable $(p-1)$-forms to the case where $0\leq p<n$, as pointed
out recently  in \cite{Kanatchikov1}, \cite{Kanatchikov2}. For
example the dissymmetry between variables $u$ and $p^\mu$ in
system (\ref{0.h}) suggests that, if the $p^\mu$'s are actually
the components of the observable $(n-1)$-form $p^\mu\omega_\mu$,
$u$ should be an observable function. Another interesting example
is the Maxwell action, where the gauge potential 1-form $A_\mu
dx^\mu$ and the Faraday $(n-2)$-form $\star dA=
\eta^{\mu\lambda}\eta^{\nu\sigma} (\partial _\mu A_\nu - \partial
_\nu A_\mu)\omega_{\lambda\sigma}$ are also ``observable'', as
proposed in \cite{Kanatchikov1}. Note that again two kinds of
approaches for defining such observable forms are possible, as in
the preceding paragraph: either our starting point is to ensure
consistency with the dynamics (this leads us in Section 3 to the
definition of OF's) or we privilege the definition which seems to
be the more appropriate for having a notion of Poisson bracket
(this leads us in Section 4 to the definition of AOF's). If we
were follow the second point of view we would be led to the
following definition, in \cite{Kanatchikov1}: a $(p-1)$-form $F$
would be observable (``Hamiltonian'' in \cite{Kanatchikov1}) if
and only if there exists a $(n-p+1)$-multivector $X_F$ such that
$dF = (-1)^{n-p+1}X_F\iN \Omega$. This definition has the
advantage that
--- thanks to a consistent definition of Lie derivatives of forms
with respect to multivectors due to W.M. Tulczyjew
\cite{Tulczyjew} --- a beautiful notion of graded Poisson bracket
between such forms can be defined, in an intrinsic way (see also
\cite{Paufler}, \cite{ForgerRomer}). These notions were used
successfully by S. Hrabak for constructing a multisymplectic
version of the Marsden--Weinstein symplectic reduction
\cite{Hrabak1} and of the BRST operator \cite{Hrabak2}.
Unfortunately such a definition of observable $(p-1)$-form would
not have nice dynamical properties. For instance if ${\cal M}:=
\Lambda^nT^{\star}(\Bbb{R}^n\times \Bbb{R})$ with $\Omega =
de\wedge \omega + dp^\mu\wedge d\phi\wedge \omega_\mu$, then the
0-form $p^1$ would be observable, since $dp^1 = (-1)^n{\partial
\over \partial \phi}\wedge {\partial \over \partial x^2}\wedge
\cdots \wedge {\partial \over \partial x^n}\iN \Omega$, but there
would be no chance for finding a law for the infinitesimal change
of $p^1$ along a curve inside a Hamiltonian $n$-curve. By that we
mean that there would be no hope for having an analogue of the
relation (\ref{0.dF}) (Corollary \ref{6.2.cor1}).\\

\noindent That is why we have tried to base ourself on the first
point of view and to choose a definition of observable
$(p-1)$-forms in order to guarantee good dynamical properties,
i.e.\,in the purpose of generalizing relation (\ref{0.dF}). A
first attempt was in \cite{HeleinKouneiher} for variational
problems concerning maps between manifolds. We propose here
another definition working for all Lepagean theories, i.e.\,more
general. Our new definition works ``collectively'', requiring to
the set of observable $(p-1)$-forms for $0\leq p<n$ that their
differentials form a sub bundle stable by exterior multiplication
and containing differentials of observable $(n-1)$-forms ({\bf
copolarization}, Section 3). This definition actually merged out
as the right notion from our efforts to generalize the dynamical
relation (\ref{0.dF}). This is the content of Theorem
\ref{6.2.thm1}.\\

\noindent Once this is done we are left with the question of
defining the bracket between an observable $(p-1)$-form $F$ and an
observable $(q-1)$-form $G$. We propose here a (partial) answer.
In Section 4 we find necessary conditions on such a bracket in
order to be consistent with the standard bracket used by
physicists in quantum field theory. Recall that this standard
bracket is built through an infinite dimensional Hamiltonian
description of fields theory. This allows us to characterize what
should be our correct bracket in two cases: either $p$ or $q$ is
equal to $n$, or $p,q\neq n$ and $p+q=n$. The second situation
arises for example for the Faraday $(n-2)$-form and the gauge
potential 1-form in electromagnetism (see Example 4'' in Section
4). However we were unable to find a general definition: this is
left as a partially open problem. Regardless, note that  this
analysis shows that the right bracket (i.e.\,from the point of
view adopted here) should have a definition which differs from
those proposed in \cite{Kanatchikov1} and also from our
previous definition in \cite{HeleinKouneiher}.\\

\noindent In Section 5 we analyze the special case where the
multisymplectic manifold is $\Lambda^nT^*{\cal N}$: this example
is important because it is the framework for Lepage--Dedecker
theory. Note that this theory has been the subject of our
companion paper \cite{HK1a}. We show that OF's and AOF's coincide
on $\Lambda^nT^*{\cal N}$. This contrasts with the de Donder--Weyl
theory in which --- like all Lepage theories obtained by a
restriction on a submanifold of $\Lambda^nT^*{\cal N}$ --- the set
of AOF's is a strict subset of the set of OF's. This singles out
the Lepage--Dedecker theory as being ``complete'': we say that
$\Lambda^nT^*{\cal N}$ is {\em pataplectic} for quoting this
property.\\

\noindent Another result in this paper is also motivated by the
important example of $\Lambda^nT^*{\cal N}$, although it may have
a larger range of application. In the Lepage--Dedecker theory
indeed the Hamiltonian function and the Hamilton equations are
invariant by deformations parallel to affine submanifolds called
{\em pseudofibers} by Dedecker \cite{Dedecker}. This sounds like
something similar to a gauge invariance but the pseudofibers may
intersect along singular sets, as already remarked by Dedecker
\cite{Dedecker}. In \cite{HK1a} we revisit this picture and
proposed an intrinsic definition of this distribution which gives
rise to a generalization  that we call the {\em generalized
pseudofiber} direction. We look here at the interplay of this
notion with observable $(n-1$)-forms, namely showing in Paragraph
4.1.3 that --- under some hypotheses --- the resulting functional
is invariant by deformation along the generalized pseudofibers
directions.\\

\noindent A last question concerns the bracket between observable
functionals obtained by integration of say $(n-1)$-forms on two
{\em different} slices. This is a crucial question if one is
concerned by the relativistic invariance of a symplectic theory.
Indeed the only way to build a relativistic invariant theory of
classical (or quantum) fields is to make sense of functionals (or
observable operators) as defined on the set of solution (each one
being a complete history in space-time), independently of the
choice of a time coordinate. This requires at least that one
should be able to define the bracket between say the observable
functionals $\int_\Sigma F$ and $\int_{\widetilde{\Sigma}} G$ even
when $\Sigma$ and ${\widetilde{\Sigma}}$ are different (imagine
they correspond to two space-like hypersurfaces). One possibility
for that is to assume that one of the two forms, say $F$ is such
that $\int_\Sigma F$ depends uniquely on the homology class of
$\Sigma$. Using Stoke's theorem one checks easily that such a
condition is possible if $\{{\cal H}, F\}= 0$. We call a {\em
dynamical} observable $(n-1)$-form any observable $(n-1)$-form
which satisfies such a relation. All that leads us to the question
of finding all such forms.\\

\noindent This problem was investigated in \cite{Kijowski1} and
discussed in \cite{GoldschmidtSternberg} (in collaboration with S.
Coleman). It led to an interesting but deceptive answer: for a
linear variational problem (i.e.\,with a linear PDE, or for free
fields) one can find a rich collection of dynamical OF's, roughly
speaking in correspondence with the set of solutions of the linear
PDE. However as soon as the problem becomes nonlinear (so for
interacting fields) the set of dynamical OF's is much more reduced
and corresponds to the symmetries of the problem (so it is in
general finite dimensional). We come back here to this question in
Section 6. We are looking at the example of a complex scalar field
with one symmetry, so that the only dynamical OF's basically
correspond to the total charge of the field. We show there that by
a kind of Noether's procedure we can enlarge the set of dynamical
OF's by including all smeared integrals of the current density.
This example illustrates the fact that gauge symmetry helps
strongly in constructing dynamical observable functionals. Another
possibility in order to enlarge the number of dynamical
functionals is when the nonlinear variational problem can be
approximated by a linear one: this gives rise to observable
functionals defined by expansions \cite{Dika},
\cite{perturb}.\\

\noindent As a conclusion we wish to insist about one of the main
motivation for multisymplectic formalisms: it is to build a
Hamiltonian theory which is consistent with the principles of
Relativity, i.e.\,being {\em covariant}. Recall for instance that
for all the multisymplectic formalisms which have been proposed
one does not need to use a privilege time coordinate. But among
them the Lepage--Dedecker is actually a quite natural framework in
order to extend this democracy between space and time coordinates
to the coordinates on fiber manifolds (i.e.\,along the fields
themselves). This is quite in the spirit of the Kaluza--Klein
theory and its modern avatars: 11-dimensional supergravity, string
theory and M-theory. Indeed in the Dedecker theory, in contrast
with the Donder--Weyl one, we do not need to split\footnote{Such a
splitting has several drawbacks, for example it causes
difficulties in order to define the stress-energy tensor.} the
variables into the horizontal (i.e.\,corresponding to space-time
coordinates) and vertical (i.e.\,non horizontal) categories. Of
course, as the reader can imagine, if we do not fix a priori the
space-time/fields splitting, many new difficulties appear as for
example: how to define forms which --- in a non covariant way of
thinking --- should be of the type $dx^\mu$, where the $x^\mu$'s
are space-time coordinates, without a space-time
background\footnote{another question which is probably related is:
how to define a ``slice'', which plays the role of a constant time
hypersurface without referring to a given space-time background ?
We propose in \cite{HK1a} a definition of such a slice which,
roughly speaking, requires a slice to be transversal to all
Hamiltonian $n$-curves, so that the dynamics only (i.e.\,the
Hamiltonian function) should determine what are the slices. We
give in \cite{HK1a} a characterization of these slices in the case
where the multisymplectic manifold is $\Lambda^nT^*{\cal N}$.} ?
One possible way is by using the (at first glance unpleasant)
definition of {\em copolarization} given in Section 3: the idea is
that forms of the ``type $dx^\mu$'' are defined collectively and
each relatively to the other ones. We believe that this notion of
copolarization corresponds somehow to the philosophy of general
relativity: the observable quantities again are not measured
directly, they are compared each to the other ones.\\

\noindent In exactly the same spirit we remark that the dynamical
law (\ref{0.dF}) can be expressed in a slightly more general form
which is: if $\Gamma$ is a Hamiltonian $n$-curve then
\begin{equation}\label{0.dFdG}
\{{\cal H}, F\}dG_{|\Gamma} = \{{\cal H}, G\}dF_{|\Gamma},
\end{equation}
for all OF's $F$ and $G$ (see Proposition \ref{3.2.2.propdyn} and
Theorem \ref{6.2.thm1}). Mathematically this is not much more
difficult than (\ref{0.dF}). However (\ref{0.dFdG}) is more
satisfactory from the point of view of relativity: no volume form
$\omega$ is singled out, the dynamics just prescribe how to
compare two observations.

\subsection{Notations}

\noindent The Kronecker symbol $\delta^{\mu}_{\nu}$ is equal to 1 if
$\mu = \nu$ and equal to 0 otherwise. We shall also set
\[
\delta^{\mu_1\cdots \mu_p}_{\nu_1\cdots \nu_p}:=
\left| \begin{array}{ccc}
\delta^{\mu_1}_{\nu_1} & \dots  & \delta^{\mu_1}_{\nu_p}\\
\vdots & & \vdots \\
\delta^{\mu_p}_{\nu_1} & \dots & \delta^{\mu_p}_{\nu_p}
\end{array}\right| .
\]
In most examples, $\eta_{\mu\nu}$ is a constant
metric tensor on $\Bbb{R}^n$ (which may be Euclidean or Minkowskian).
The metric on his dual space his $\eta^{\mu\nu}$. Also, $\omega$ will
often denote a volume form on some space-time: in local coordinates
$\omega=dx^1\wedge \cdots \wedge dx^n$ and we will use several times the notation
$\omega_\mu:= {\partial \over \partial x^\mu}\iN \omega$,
$\omega_{\mu\nu}:= {\partial \over \partial x^\mu}\wedge
{\partial \over \partial x^\nu}\iN \omega$, etc. Partial derivatives
${\partial \over \partial x^\mu}$ and ${\partial \over \partial p_{\alpha_1\cdots \alpha_n}}$
will be sometime abbreviated by $\partial _\mu$ and $\partial ^{\alpha_1\cdots \alpha_n}$
respectively.\\

\noindent
When an index or a symbol is omitted in the middle of a
sequence of indices or symbols, we denote this omission by $\widehat{\,}$.
For example $a_{i_1\cdots \widehat{i_p}\cdots i_n}:=
a_{i_1\cdots i_{p-1}i_{p+1}\cdots i_n}$,
$dx^{\alpha_1}\wedge \cdots \wedge \widehat{dx^{\alpha_\mu}}\wedge \cdots \wedge dx^{\alpha_n} :=
dx^{\alpha_1}\wedge \cdots \wedge dx^{\alpha_{\mu-1}}\wedge dx^{\alpha_{\mu+1}}\wedge \cdots \wedge dx^{\alpha_n}$.\\

\noindent If ${\cal N}$ is a manifold
and ${\cal FN}$ a fiber bundle over
${\cal N}$, we denote by $\Gamma({\cal N},{\cal FN})$ the set of smooth sections of
${\cal FN}$. Lastly we use the notations concerning the exterior algebra of multivectors and
differential forms, following W.M.\,Tulczyjew \cite{Tulczyjew}.
If ${\cal N}$ is a differential $N$-dimensional manifold and $0\leq k\leq N$,
$\Lambda^kT{\cal N}$ is the bundle over ${\cal N}$ of $k$-multivectors
($k$-vectors in short)
and $\Lambda^kT^{\star}{\cal N}$ is the bundle of differential forms of degree $k$
($k$-forms in short). Setting $\Lambda T{\cal N}:= \oplus_{k=0}^N\Lambda^kT{\cal N}$
and $\Lambda T^{\star}{\cal N}:= \oplus_{k=0}^N\Lambda^kT^{\star}{\cal N}$, there exists
a unique duality evaluation map between $\Lambda T{\cal N}$ and $\Lambda T^{\star}{\cal N}$
such that for every decomposable $k$-vector field $X$, i.e.\,of the form
$X=X_1\wedge \cdots \wedge X_k$, and for every $l$-form $\mu$, then
$\langle X,\mu\rangle = \mu(X_1,\cdots ,X_k)$ if $k=l$ and $=0$ otherwise.
Then interior products $\iN $ and $\Ni$ are operations
defined as follows. If $k\leq l$, the product
$\iN :\Gamma({\cal N},\Lambda^kT{\cal N})\times \Gamma({\cal N},\Lambda^lT^{\star}{\cal N})\longrightarrow
\Gamma({\cal N},\Lambda^{l-k}T^{\star}{\cal N})$
is given by
$$\langle Y,X\iN \mu\rangle = \langle X\wedge Y,\mu\rangle ,\quad
\forall (l-k)\hbox{-vector }Y.$$
And if $k\geq l$, the product $\Ni :\Gamma({\cal N},\Lambda^kT{\cal N})\times
\Gamma({\cal N},\Lambda^lT^{\star}{\cal N})\longrightarrow
\Gamma({\cal N},\Lambda^{k-l}T{\cal N})$ is given by
\[
\langle X\Ni \mu,\nu\rangle = \langle X,\mu\wedge \nu\rangle,\quad
\forall (k-l)\hbox{-form }\nu.
\]

\section{Basic facts about multisymplectic manifolds}
We recall here the general framework introduced in \cite{HK1a}.
\subsection{Multisymplectic manifolds}

\begin{defi}\label{2.1.def1}
Let ${\cal M}$ be a differential manifold. Let $n\in \Bbb{N}$ be
some positive integer. A smooth $(n+1)$-form $\Omega$ on ${\cal
M}$ is a {\bf multisymplectic} form if and only if
\begin{enumerate}
\item [(i)] $\Omega$ is non degenerate, i.e.\,$\forall m\in {\cal
M}$, $\forall \xi \in T_m{\cal M}$, if $\xi \iN \Omega_m = 0$,
then $\xi = 0$ \item [(ii)] $\Omega$ is closed, i.e.\,$d\Omega =
0$.
\end{enumerate}
Any manifold ${\cal M}$ equipped with a multisymplectic form
$\Omega$ will be called a {\bf multisymplectic} manifold.
\end{defi}
In the following, $N$ denotes the dimension of ${\cal M}$. For any
$m\in {\cal M}$ we define the set
\[
D^n_m{\cal M}:= \{X_1\wedge \cdots \wedge X_n\in \Lambda^nT_m{\cal M}/
X_1,\cdots , X_n\in T_m{\cal M}\},
\]
of {\bf decomposable} $n$-vectors and denote by $D^n{\cal M}$ the
associated bundle.

\begin{defi}\label{2.1.def2}
Let ${\cal H}$ be a smooth real valued function defined over a
multisymplectic manifold $({\cal M}, \Omega)$. A Hamiltonian
$n$-curve $\Gamma$ is a $n$-dimensional submanifold of ${\cal M}$
such that for any $m\in \Gamma$, there exists a $n$-vector $X$ in
$\Lambda^nT_m\Gamma$ which satisfies
$$X\iN \Omega = (-1)^nd{\cal H}.$$
We denote by ${\cal E}^{\cal H}$ the set of all such Hamiltonian $n$-curves. We
also write for all $m\in {\cal M}$, $[X]^{\cal H}_m:=\{X\in D^n_m{\cal M}/X\iN \Omega = (-1)^nd{\cal H}_m\}$.
\end{defi}

\noindent {\bf Example 1} --- {\em The Lepage--Dedecker
multisymplectic manifold $(\Lambda^nT^*{\cal N}, \Omega)$} --- It
was studied in \cite{HK1a}. Here $\Omega:=d\theta$ where $\theta$
is the generalized Poincar\'e--Cartan 1-form defined by
$\theta(X_1,\cdots X_n)=\langle \Pi^*X_1,\cdots
,\Pi^*X_n,p\rangle$, $\forall X_1,\cdots,X_n\in
T_{(q,p)}(\Lambda^nT^*{\cal N})$ and $\Pi:\Lambda^nT^*{\cal
N}\longrightarrow {\cal N}$ is the canonical projection. If we use
local coordinates $\left(q^\alpha\right)_{1\leq \alpha\leq n+k}$
on ${\cal N}$, then a basis of $\Lambda^nT^*_q{\cal N}$ is the
family $\left( dq^{\alpha_1}\wedge \cdots \wedge
dq^{\alpha_n}\right)_{1\leq \alpha_1< \cdots <\alpha_n \leq n+k}$
and we denote by $p_{\alpha_1\cdots \alpha_n}$ the coordinates on
$\Lambda^nT^*_q{\cal N}$ in this basis. Then $\Omega$ writes
\begin{equation}\label{2.2.2.theta}
\Omega:= \sum_{1\leq \alpha_1<\cdots <\alpha_n\leq n+k}
dp_{\alpha_1\cdots \alpha_n}\wedge dq^{\alpha_1}\wedge \cdots
\wedge dq^{\alpha_n}.
\end{equation}
If the Hamiltonian function ${\cal H}$ is associated to a
Lagrangian variational problem on $n$-dimensional submanifolds of
${\cal N}$ by means of a Legendre correspondence (see \cite{HK1a},
\cite{Dedecker}) we then say that ${\cal H}$ is a {\bf Legendre image Hamiltonian function}.\\

\noindent A particular case is when ${\cal N} = {\cal X}\times
{\cal Y}$ where ${\cal X}$ and ${\cal Y}$ are manifolds of
dimension $n$ and $k$ respectively. This situation occurs when we
look at variational problems on maps $u:{\cal X}\longrightarrow
{\cal Y}$. We denote by $q^\mu=x^\mu$, if $1\leq \mu\leq n$,
coordinates on ${\cal X}$ and by $q^{n+i}=y^i$, if $1\leq i\leq
k$, coordinates on ${\cal Y}$. We also denote by $e:=p_{1\cdots
n}$, $p^{\mu}_i:=p_{1\cdots (\mu-1)i(\mu+1)\cdots n}$,
$p^{\mu_1\mu_2}_{i_1i_2}:=p_{1\cdots (\mu_1-1)i_1(\mu_1+1)\cdots
(\mu_2-1)i_2(\mu_2+1)\cdots n}$, etc., so that
\[
\Omega=de\wedge \omega + \sum_{j=1}^n\sum_{\mu_1<\cdots <\mu_j}
\sum_{i_1<\cdots <i_j}dp^{\mu_1\cdots \mu_j}_{i_1\cdots i_j}\wedge
\omega_{\mu_1\cdots \mu_j}^{i_1\cdots i_j},
\]
where, for $1\leq p\leq n$,
\[
\begin{array}{ccl}
\omega & := & dx^1\wedge \cdots  \wedge dx^n\\
\omega^{i_1\cdots i_p}_{\mu_1\cdots \mu_p} & := & dy^{i_1}\wedge \cdots  \wedge
dy^{i_p}\wedge \left( {\partial \over \partial x^{\mu_1}}\wedge \cdots  \wedge
{\partial \over \partial x^{\mu_p}}\iN \omega\right) .
\end{array}
\]
Note that if ${\cal H}$ is the Legendre image of a Lagrangian
action of the form $\int_{\cal X} \ell (x,u(x),du(x))\omega$ and
if we denote by $p^*$ all coordinates $p^{\mu_1\cdots
\mu_j}_{i_1\cdots i_j}$ for $j\geq 1$, we can always write ${\cal
H}(q,e,p^*) = e + H(q,p^*)$ (see for instance \cite{Dedecker},
\cite{HeleinKouneiher},\cite{HK1a}).\\

\noindent Other examples are provided by considering the
restriction of $\Omega$ on any smooth submanifold of
$\Lambda^nT^*{\cal N}$,
like for instance the following.\\

\noindent
{\bf Example 2} --- {\em The de Donder--Weyl manifold ${\cal M}^{dDW}_q$} ---
It is the submanifold of
$\Lambda^nT^*_q{\cal N}$ defined by the constraints
$p^{\mu_1\cdots \mu_j}_{i_1\cdots i_j} = 0$, for all $j\geq 2$.
We thus have
\[
\Omega^{dDW}=de\wedge \omega + \sum_\mu \sum_idp^\mu_i\wedge \omega_\mu^i.
\]

\noindent {\bf Example 3} --- {\em The Palatini formulation of
pure gravity in 4-dimensional space-time} --- (see also
\cite{Rovelli}) We describe here the Riemannian (non Minkowskian)
version of it. We consider $\Bbb{R}^4$ equipped with its standard
metric $\eta_{IJ}$ and with the standard volume 4-form
$\epsilon_{IJKL}$. Let $\mathfrak{p}\simeq \left\{(a,v)\simeq
\left(\begin{array}{cc}a & v\\ 0 & 1\end{array}\right)/ a\in
so(4),v\in \Bbb{R}^4\right\}$ $\simeq so(4)\ltimes \Bbb{R}^4$ be
the Lie algebra of the Poincar\'e group acting on $\Bbb{R}^4$. Now
let ${\cal X}$ be a 4-dimensional manifold, the ``space-time'',
and consider ${\cal M}:=\mathfrak{p}\otimes T^*{\cal X}$, the
fiber bundle over ${\cal X}$ of 1-forms with coefficients in
$\mathfrak{p}$. We denote by $(x,e,A)$ a point in ${\cal M}$,
where $x\in {\cal X}$, $e\in \Bbb{R}^4\otimes T^*_x$ and $A\in
so(4)\otimes T^*_x$. We shall work is the open subset of ${\cal
M}$ where $e$ is rank 4 (so that the 4 components of $e$ define a
coframe on $T_x{\cal X}$). First using the canonical projection
$\Pi:{\cal M}\longrightarrow {\cal X}$ one can define a
$\mathfrak{p}$-valued 1-form $\theta^{\mathfrak{p}}$ on ${\cal M}$
(similar to the Poincar\'e--Cartan 1-form) by
\[
\forall (x,e,A)\in {\cal M}, \forall X\in T_{(x,e,A)}{\cal X}, \quad
\theta^{\mathfrak{p}}_{(x,e,A)}(X) := (e(\Pi^*X),A(\Pi^*X)).
\]
Denoting (for $1\leq I,J\leq 4$) by $T^I:\mathfrak{p}\longrightarrow \Bbb{R}$,
$(a,v)\longmapsto v^I$ and by
$R^I_J:\mathfrak{p}\longrightarrow \Bbb{R}$, $(a,v)\longmapsto a^I_J$,
the coordinate mappings we can define a 4-form on ${\cal M}$ by
\[
\theta_{Palatini}:= {1\over 4!}\epsilon_{IJKL}\eta^{LN}(T^I\circ \theta^{\mathfrak{p}}) \wedge
(T^J\circ \theta^{\mathfrak{p}})\wedge \left(R^K_N\circ d\theta^{\mathfrak{p}}+
(R^K_M\circ \theta^{\mathfrak{p}}) \wedge (R^M_N\circ \theta^{\mathfrak{p}})\right).
\]
Now consider any section of ${\cal M}$
over ${\cal X}$. Write it as $\Gamma:=\{(x,e_x,A_x)/x\in {\cal X}\}$ where now
$e$ and $A$ are {\em 1-forms} on $x$ (and not coordinates anymore). Then
\[
\int_\Gamma \theta_{Palatini} = \int_{\cal X}{1\over 4!}\epsilon_{IJKL}\eta^{LN}
e^I\wedge e^J\wedge F^K_L,
\]
where $F^I_J:= dA^I_J + A^I_K\wedge A^K_J$ is the curvature of the connection 1-form $A$. We recognize the
Palatini action for pure gravity in 4 dimensions: this functional has the property
that a critical point of it provides us with
a solution of Einstein gravity equation $R_{\mu\nu} -{1\over 2}g_{\mu\nu} = 0$ by setting
$g_{\mu\nu}:= \eta_{IJ}e^I_\mu e^J_\nu$. By following the same steps as in the proof of Theorem
2.2 in \cite{HK1a} one proves that a 4-dimensional submanifold $\Gamma$ which is a critical point of this action,
satisfies the Hamilton equation $X\iN \Omega_{Palatini} = 0$, where $\Omega_{Palatini}:= d\theta_{Palatini}$. Thus
$({\cal M},\Omega_{Palatini})$ is a multisymplectic manifold naturally associated to gravitation.
In the above construction, by replacing $A$ and $F$ by their self-dual parts $A_+$ and
$F_+$ (and so reducing the gauge group to $SO(3)$) one obtains the Ashtekar action.\\
Remark also that a similar construction can be done for the Chern--Simon action in dimension 3.
\begin{defi}\label{2.1.def3}
A {\bf symplectomorphism} $\phi$ of a multisymplectic manifold
$({\cal M},\Omega)$ is a smooth diffeomorphism $\phi:{\cal
M}\longrightarrow {\cal M}$ such that $\phi^*\Omega = \Omega$. An
{\bf infinitesimal symplectomorphism} is a vector field $\xi\in
\Gamma({\cal M},T{\cal M})$ such that $L_\xi\Omega =0$. We denote
by $\mathfrak{sp}_0{\cal M}$ the set of infinitesimal
symplectomorphisms of $({\cal M}, \Omega)$.
\end{defi}
Note that, since $\Omega$ is closed, $L_\xi\Omega=d(\xi\iN \Omega)$, so that a vector field
$\xi$ belongs to $\mathfrak{sp}_0{\cal M}$ if and only if $d(\xi\iN \Omega)=0$. Hence
if the homology group $H^n({\cal M})$ is trivial there exists an $(n-1)$-form $F$ on
${\cal M}$ such that $dF + \xi\iN \Omega = 0$: such an $F$ will be called an algebraic observable $(n-1)$-form
(see Section 3.3).

\subsection{Pseudofibers}
It may happen that the dynamical structure encoded by the data of
a multisymplectic manifold $({\cal M}, \Omega)$ and a Hamiltonian
function ${\cal H}$ is invariant by deformations along some
particular submanifolds called {\bf pseudofibers}. This situation
is similar to gauge theory where two fields which are equivalent
through a gauge transformation are supposed to correspond to the
same physical state. A slight difference however lies in the fact
that pseudofibers are not fibers in general and can intersect
singularly. This arises for instance when ${\cal M} =
\Lambda^nT^*{\cal N}$ and ${\cal H}$ is a Legendre image Hamiltonian function (see
\cite{Dedecker}, \cite{HK1a}). In the latter situation
the singular intersections of pseudofibers picture
geometrically the constraints caused by gauge invariance. All that
is the origin of the following definitions.

\begin{defi}\label{3.3.4.gpd}
For all Hamiltonian function ${\cal H}:{\cal M}\longrightarrow
\Bbb{R}$ and for all $m\in {\cal M}$ we define the {\bf
generalized pseudofiber direction} to be
\begin{equation}\label{3.3.4.lh}
\begin{array}{ccl}
L^{\cal H}_m & := & \{\xi \in T_m{\cal M}/\forall X\in [X]^{\cal H}_m,
\forall \delta X\in T_XD^n_m{\cal M}, \xi\iN \Omega(\delta X) = 0\}\\
 & = & \displaystyle \left(T_{[X]_m^{\cal H}}D^n_m{\cal M}\iN \Omega\right)^{\perp}.
\end{array}
\end{equation}
And we write $L^{\cal H}:=\cup_{m\in {\cal M}}L^{\cal H}_m\subset T{\cal M}$ for
the associated bundle.
\end{defi}
Note that in the case where ${\cal M} = \Lambda^nT^*{\cal N}$ and
${\cal H}$ is the Legendre image Hamiltonian (see Section 2.1)
then all generalized pseudofibers directions $L^{\cal H}_m$ are
``vertical'' i.e.\,$L^{\cal H}_m\subset \hbox{Ker}d\Pi_m \simeq
\Lambda^n_{\Pi(m)}T^*{\cal N}$, where $\Pi: \Lambda^nT^*{\cal
N}\longrightarrow {\cal N}$.
\begin{defi}\label{3.3.4.def}
We say that ${\cal H}$ is {\bf pataplectic invariant} if
\begin{itemize}
\item $\forall \xi\in L^{\cal H}_m$, $d{\cal H}_m(\xi)=0$
\item for all Hamiltonian $n$-curve $\Gamma\in {\cal E}^{\cal H}$, for all
vector field $\xi$ which is a smooth section of $L^{\cal H}$, then,
for $s\in \Bbb{R}$ sufficiently small, $\Gamma_s:=e^{s\xi}(\Gamma)$ is also
a Hamiltonian $n$-curve.
\end{itemize}
\end{defi}
We proved in \cite{HK1a} that, if ${\cal M}$ is an open subset of
$\Lambda^nT^*{\cal N}$, any function on ${\cal M}$ which is a
Legendre image Hamiltonian is pataplectic invariant.

\subsection{Functionals defined by means of integrations of forms}
An example of functional on ${\cal E}^{\cal H}$ is obtained
by choosing a codimension $r$ submanifold $\Sigma$ of ${\cal M}$ (for $1\leq r\leq n$)
and a $(p-1)$-form $F$ on ${\cal M}$ with $r = n-p+1$: it leads to the definition of
\[
\begin{array}{cccl}
\displaystyle \int_\Sigma F : & \displaystyle {\cal E}^{\cal H} & \longrightarrow & \R\\
 & \Gamma & \longmapsto  &  \displaystyle \int_{\Sigma\cap \Gamma}F
\end{array}
\]
But in order for this definition to be meaningful one should first
make sure that the intersection $\Sigma\cap \Gamma$ is a
$(p-1)$-dimensional submanifold. This is true if $\Sigma$ fulfills
the following definition. (See also \cite{HK1a}.)
\begin{defi}\label{2.1.def40}
Let ${\cal H}$ be a smooth real valued function defined over a
multisymplectic manifold $({\cal M}, \Omega)$. A {\bf slice of
codimension} $r$ is a cooriented submanifold $\Sigma$ of ${\cal
M}$ of codimension $r$ such that for any $\Gamma\in {\cal E}^{\cal
H}$, $\Sigma$ is transverse to $\Gamma$. By {\bf cooriented} we
mean that for each $m\in \Sigma$, the quotient space $T_m{\cal
M}/T_m\Sigma$ is oriented continuously in function of $m$.
\end{defi}
If we represent such a submanifold as a level set of a given
function into $\R^r$ then it suffices that the restriction of such
a function on any Hamiltonian $n$-curve have no critical point. We
then say that the function is {\em $r$-regular}. In \cite{HK1a} we
give a characterization of $r$-regular functions in
$\Lambda^nT^*{\cal N}$.\\

\noindent A second question concerns then the choice of $F$: what
are the conditions on $F$ for $\int_\Sigma F$ to be a physically
observable functional ? Clearly the answer should agree with the
experience of physicists, i.e.\,be based in the knowledge of all
functionals which are physically meaningful. Our aim is here to
understand which mathematical properties would characterize all
such functionals. In the following we explore this question, by
following two possible points of view.

\section{The ``dynamical'' point of view}

\subsection{Observable $(n-1)$-forms}
We define here the concept of {\em observable $(n-1)$-forms} $F$.
The idea is that given a point $m\in {\cal M}$
and a Hamiltonian function ${\cal H}$, if $X(m)\in [X]^{\cal H}_m$, then
$\langle X(m),dF_m\rangle$
should not depend on the choice of $X(m)$ but only on $d{\cal H}_m$.

\subsubsection{Definitions}
\begin{defi}\label{3.2.1.def}
Let $m\in {\cal M}$ and $a\in \Lambda^nT^{\star}_m{\cal M}$; $a$
is called a {\bf copolar} $n$-form if and only if there exists an
open dense subset ${\cal O}^a_m{\cal M}\subset D^n_m{\cal M}$ such
that
\begin{equation}\label{dynaobs}
\forall X,\tilde{X}\in {\cal O}^a_m{\cal M},\quad
X\iN \Omega = \tilde{X}\iN \Omega\ \Longrightarrow \ a(X) = a(\tilde{X}).
\end{equation}
We denote by $P^n_mT^{\star}{\cal M}$ the set of copolar $n$-forms
at $m$. A $(n-1)$-form $F$ on ${\cal M}$ is called {\bf
observable} if and only if for every $m\in {\cal M}$, $dF_m$ is
copolar i.e.\,$dF_m\in P^n_mT^{\star}{\cal M}$. We denote by
$\mathfrak{P}^{n-1}{\cal M}$ the set of observable $(n-1)$-forms
on ${\cal M}$.
\end{defi}
{\bf Remark} ---
For any $m\in {\cal M}$, $P^nT^{\star}_m{\cal M}$ is a vector space (in particular
if $a,b\in P^nT^{\star}_m{\cal M}$ and $\lambda,\mu\in \Bbb{R}$ then $\lambda a + \nu b\in
P^nT^{\star}_m{\cal M}$ and we can choose ${\cal O}^{\lambda a + \nu b}_m{\cal M}
= {\cal O}^a_m{\cal M}\cap {\cal O}^b_m{\cal M}$) and so
it is possible to construct a basis $(a_1,\cdots ,a_r)$ for this  space. Hence for any
$a\in P^nT^{\star}_m{\cal M}$ we can write $a = t^1a_1+\cdots +t^ra_r$ which
implies that we can choose ${\cal O}^a_m{\cal M} = \cap_{s=1}^r{\cal O}^{a_s}_m{\cal M}$.
So having choosing such a basis $(a_1,\cdots ,a_r)$ we will denote by
${\cal O}_m{\cal M}:=\cap_{s=1}^r{\cal O}^{a_s}_m{\cal M}$ (it is still open and dense in $D^n_m{\cal M}$)
and in the following we will replace ${\cal O}^a_m{\cal M}$ by ${\cal O}_m{\cal M}$ in the above definition.
We will also denote by ${\cal O}{\cal M}$ the associated bundle.

\begin{lemm}\label{3.2.1.lemme}
Let $\phi:{\cal M}\longrightarrow {\cal M}$ be a symplectomorphism and
$F\in \mathfrak{P}^{n-1}{\cal M}$. Then $\phi^*F\in \mathfrak{P}^{n-1}{\cal M}$.
As a corollary, if $\xi\in \mathfrak{sp}_0{\cal M}$ (i.e.\,is an infinitesimal symplectomorphism)
and $F\in \mathfrak{P}^{n-1}{\cal M}$, then $L_\xi F\in \mathfrak{P}^{n-1}{\cal M}$.
\end{lemm}
{\em Proof} --- For any $n$-vector fields $X$ and $\widetilde{X}$, which are sections
of ${\cal OM}$, and for any $F\in \mathfrak{P}^{n-1}{\cal M}$,
\[
X\iN \Omega = \widetilde{X}\iN \Omega \Longleftrightarrow
X\iN \phi^*\Omega = \widetilde{X}\iN \phi^*\Omega \Longleftrightarrow
(\phi_*X)\iN \Omega = (\phi_*\widetilde{X})\iN \Omega
\]
implies
\[
dF(\phi_*X) = dF(\phi_*\widetilde{X}) \Longleftrightarrow
\phi_*dF(X) = \phi_*dF(\widetilde{X}) \Longleftrightarrow
d(\phi_*F)(X) = d(\phi_*F)(\widetilde{X}).
\]
Hence $\phi_*F\in \mathfrak{P}^{n-1}{\cal M}$. \bbox

\noindent Assume that a given Hamiltonian function ${\cal H}$ on
${\cal M}$ is such that $[X]^{\cal H}_m\subset {\cal O}_m{\cal
M}$. Then we shall say that ${\cal H}$ is {\bf admissible}. If
${\cal H}$ is so, we define the {\bf pseudobracket} for all
observable $(n-1)$-form $F\in \mathfrak{P}^{n-1}{\cal M}$
$$\{{\cal H},F\} := X\iN dF = dF(X),$$
where $X$ is any $n$-vector in $[X]^{\cal H}_m$. Remark that,
using the same notations as in Example 1, if ${\cal M} =
\Lambda^nT^*({\cal X}\times {\cal Y})$ and ${\cal H}(x,u,e,p^*) =
e + H(x,u,p^*)$, then $\{{\cal H},x^1dx^2\wedge \cdots \wedge
dx^n\} = 1$.

\subsubsection{Dynamics equation using pseudobrackets}
Our purpose here is to generalize the classical well-known relation $dF/dt = \{H,F\}$ of the classical
mechanics.
\begin{prop}\label{3.2.2.propdyn}
Let ${\cal H}$ be a smooth admissible Hamiltonian on ${\cal M}$ and $F$, $G$ two observable
$(n-1)$-forms with ${\cal H}$. Then $\forall \Gamma\in {\cal E}^{\cal H}$,
$$\{{\cal H},F\}dG_{|\Gamma} = \{{\cal H},G\}dF_{|\Gamma}.$$
\end{prop}
{\em Proof} --- This result is equivalent to proving that, if $X\in D_m^n{\cal M}$
is different of 0 and is tangent to $\Gamma$ at $m$, then
\begin{equation}\label{3.2.1.FdG}
\{{\cal H},F\}dG(X) = \{{\cal H},G\}dF(X).
\end{equation}
Note that by rescaling, we can assume w.l.g.\,that $X\iN \Omega = (-1)^nd{\cal H}$,
i.e.\,$X\in [X]^{\cal H}_m$. But then (\ref{3.2.1.FdG}) is equivalent to the obvious relation
$\{{\cal H},F\}\{{\cal H},G\} = \{{\cal H},G\}\{{\cal H},F\}$. \bbox

\noindent This result immediately implies the following result.
\begin{coro}\label{3.2.2.corodyn}
Let ${\cal H}$ be a smooth admissible Hamiltonian function on
${\cal M}$. Assume that $F$ and $G$ are observable $(n-1)$-forms
with ${\cal H}$ and that $\{{\cal H},G\} = 1$ (see the remark at
the end of Paragraph 3.1.1). Then denoting $\omega:=dG$ we have:
$$\forall \Gamma\in {\cal E}^{\cal H},\quad \{{\cal H},F\}\omega_{|\Gamma} = dF_{|\Gamma}.$$
\end{coro}

\subsection{Observable $(p-1)$-forms}
We now introduce
observable $(p-1)$-forms, for $1\leq p< n$. The simplest situation where
such forms play some role occurs when studying variational problems on
maps $u:{\cal X}\longrightarrow {\cal Y}$: any coordinate function $y^i$ on ${\cal Y}$
is an observable functional, which at least in a classical context can be measured.
This observable 0-form can be considered as canonically conjugate with the
momentum observable form $\partial /\partial y^i\iN \theta$.
A more complex situation is given by Maxwell equations: as proposed for the first
time by I.\,Kanatchikov in \cite{Kanatchikov1} (see also \cite{HeleinKouneiher}),
the electromagnetic gauge potential and the Faraday fields can be modelled in
an elegant way by observable 1-forms and $(n-2)$-forms respectively.\\

\noindent
{\bf Example 4} --- {\em Maxwell equations on Minkowski space-time} ---
Assume here for simplicity that ${\cal X}$ is
the four-dimensional Minkowski space. Then the gauge field is a 1-form
$A(x) = A_{\mu}(x)dx^{\mu}$
defined over ${\cal X}$, i.e.\,a section of the bundle $T^{\star}{\cal X}$.
The action functional in the presence of
a (quadrivector) current field $j(x)= j^{\mu}(x)\partial /\partial x^{\mu}$ is
$\int_{\cal X}l(x,A,dA)\omega$, where
$\omega = dx^0\wedge dx^1\wedge dx^2\wedge dx^3$ and
$$l(x,A,dA) = - {1\over 4}F_{\mu\nu}F^{\mu\nu} - j^{\mu}(x)A_{\mu},$$
where $F_{\mu\nu}:= \partial _{\mu}A_{\nu} - \partial _{\nu}A_{\mu}$ and
$F^{\mu\nu}:= \eta^{\mu\lambda}\eta^{\nu\sigma}F_{\lambda\sigma}$ (see \cite{HeleinKouneiher}).
The associated multisymplectic
manifold is then ${\cal M}:= \Lambda^4T^{\star}(T^{\star}{\cal X})$ with the multisymplectic form
$$\Omega = de\wedge \omega + \sum_{\mu,\nu}dp^{A_{\mu}\nu}\wedge
da_{\mu}\wedge \omega_\nu + \cdots $$
For simplicity we restrict ourself to the de Donder--Weyl submanifold (where all momentum
coordinates excepted $e$ and $p^{A_{\mu}\nu}$ are set to 0). This implies automatically
the further constraints $p^{A_{\mu}\nu} + p^{A_{\nu}\mu} = 0$, because the Legendre
correspondence degenerates when restricted to the de Donder--Weyl submanifold.
We shall hence denote
\[
p^{\mu\nu}:= p^{A_{\mu}\nu} = - p^{A_{\nu}\mu}.
\]
Let us call ${\cal M}^{Max}$ the resulting multisymplectic manifold.
Then the multisymplectic form can be written as
\[
\Omega = de\wedge \omega + d\pi\wedge da\quad \hbox{where }
a:=a_\mu dx^\mu\hbox{ and }
\pi:= -{1\over 2}\sum_{\mu,\nu}p^{\mu\nu}\omega_{\mu\nu}.
\]
(We also have $d\pi\wedge da = \sum_{\mu,\nu}dp^{\mu\nu}\wedge da_\mu\wedge \omega_\nu$.)
Note that here $a_{\mu}$ is not anymore a function of $x$ but a fiber coordinate.
The Hamiltonian is then
\[
{\cal H}(x,a,p) = e - {1\over 4}
\eta_{\mu\lambda}\eta_{\nu\sigma}p^{\mu\nu}p^{\lambda\sigma}
+ j^{\mu}(x)a_{\mu}.
\]

\subsubsection{Copolarization and polarization}
The dynamical properties of $(p-1)$-forms are more subtle for
$1\leq p<n$ than for $p=n$, since if $F$ is such a $(p-1)$-form
then there is no way a priori to ``evaluate'' $dF$ along a
Hamiltonian $n$-vector $X$ and a fortiori no way to make sense
that ``$dF_{|X}$ should not depend on $X$ but on $d{\cal H}_m$''.
This situation is in some sense connected with the problem of
measuring a distance in relativity: we actually never measure the
distance between two points (finitely or infinitely close) but we
do {\em compare} observable quantities (distance, time) between
themselves. This analogy suggests us the conclusion that we should
define observable $(p-1)$-forms {\em collectively}. The idea is
naively that if for instance $F_1,\cdots ,F_n$ are 0-forms, then
they are observable forms if $dF_1\wedge \cdots \wedge dF_n$ can
be ``evaluated'' in the sense that $dF_1\wedge \cdots \wedge
dF_n(X)$ does not depend on the choice of the Hamiltonian
$n$-vector $X$ but on $d{\cal H}$. So it just means that
$dF_1\wedge \cdots \wedge dF_n$ is copolar. We keeping this in
mind in the following definitions.

\begin{defi}\label{6.1.def4}
Let ${\cal M}$ be a multisymplectic manifold. A {\em
copolarization} on ${\cal M}$ is a smooth vector subbundle denoted
by $P^*T^{\star}{\cal M}$ of $\Lambda^*T^{\star}{\cal M}$
satisfying the following properties
\begin{itemize}
\item $P^*T^{\star}{\cal M}:=\oplus_{j=1}^NP^jT^{\star}{\cal M}$, where
$P^jT^{\star}{\cal M}$ is a subbundle of $\Lambda^jT^{\star}{\cal M}$
\item for each $m\in {\cal M}$, $(P^*T^{\star}_m{\cal M},+,\wedge )$ is a subalgebra of
$(\Lambda^*T^{\star}_m{\cal M},+,\wedge )$
\item $\forall m\in {\cal M}$ and $\forall a\in \Lambda^nT^{\star}_m{\cal M}$,
$a\in P^nT^{\star}_m{\cal M}$ if and only if $\forall X,\widetilde{X}\in {\cal O}_m$,
$X\iN \Omega = \widetilde{X}\iN \Omega \Longrightarrow a(X) = a(\widetilde{X})$.
\end{itemize}
\end{defi}

\begin{defi}\label{6.1.def5}
Let ${\cal M}$ be a multisymplectic manifold with a copolarization
$P^*T^{\star}{\cal M}$. Then for $1\leq p\leq n$, the set of {\em
observable $(p-1)$-forms} associated to $P^*T^{\star}{\cal M}$ is
the set of smooth $(p-1)$-forms $F$ (sections of
$\Lambda^{p-1}T^{\star}{\cal M}$) such that for any $m\in {\cal
M}$, $dF_m\in P^pT^{\star}_m{\cal M}$. This set is denoted by
$\mathfrak{P}^{p-1}{\cal M}$. We shall write $\mathfrak{P}^*{\cal
M}:= \oplus_{p=1}^n\mathfrak{P}^{p-1}{\cal M}$.
\end{defi}

\begin{defi}\label{6.1.def6}
Let ${\cal M}$ be a multisymplectic manifold with a copolarization
$P^*T^{\star}{\cal M}$. For each $m\in {\cal M}$ and $1\leq p\leq
n$, consider the equivalence relation in $\Lambda^pT_m{\cal M}$
defined by $X \sim \widetilde{X}$ if and only if $\langle
X,a\rangle = \langle \widetilde{X},a\rangle$, $\forall a\in
P^pT^{\star}_m{\cal M}$. Then the quotient set $P^pT_m{\cal
M}:=\Lambda^pT_m{\cal M}/\sim$ is called a {\bf polarization} of
${\cal M}$.
If $X\in \Lambda^pT_m{\cal M}$, we denote by $[X]\in P^pT_m{\cal M}$ its equivalence class.\\
Equivalently a {\bf polarization} can be defined as being the
dual bundle of the copolarization $P^*T^{\star}{\cal M}$.
\end{defi}

\subsubsection{Examples of copolarization}
On an open subset ${\cal M}$ of $\Lambda^nT^*{\cal N}$ we can
construct the following copolarization, that we will call {\em
standard}: for each $(q,p)\in \Lambda^nT^*{\cal N}$ and for $1\leq
p\leq n-1$ we take $P^p_{(q,p)}T^*{\cal M}$ to be the vector space
spanned by $\left(dq^{\alpha_1}\wedge \cdots \wedge dq^{\alpha_p}
\right)_{1\leq \alpha_1<\cdots <\alpha_p\leq n+k}$; and
$P^n_{(q,p)}T^*{\cal M}=\{\xi\iN \Omega/\xi\in T_m{\cal M}\}$. It
means that $P^n_{(q,p)}T^*{\cal M}$ contains all
$dq^{\alpha_1}\wedge \cdots \wedge dq^{\alpha_n}$'s plus forms of
the type $\xi\iN \Omega$, for $\xi\in T_q{\cal N}$ (which
corresponds to differentials of momentum and energy-momentum observable $(n-1)$-forms).\\

\noindent
Another situation is the following.\\

\noindent {\bf Example 4'} --- {\em Maxwell equations} --- We
continue Example 4 given at the beginning of this Section. In
${\cal M}^{Max}$ with the multisymplectic form $\Omega = de\wedge
\omega + d\pi\wedge da$ the more natural choice of copolarization
is:
\begin{itemize}
\item $\displaystyle P^1_{(q,p)}T^*{\cal M}^{Max} = \bigoplus_{0\leq \mu\leq 3}\Bbb{R}dx^\mu$.
\item $\displaystyle P^2_{(q,p)}T^*{\cal M}^{Max} = \bigoplus_{0\leq \mu_1<\mu_2\leq 3}\Bbb{R}
dx^{\mu_1}\wedge dx^{\mu_2}\oplus \Bbb{R}da$, where $da:=\sum_{\mu=0}^3da_\mu\wedge dx^\mu$.
\item $\displaystyle P^3_{(q,p)}T^*{\cal M}^{Max} = \bigoplus_{0\leq \mu_1<\mu_2<\mu_3\leq 3}\Bbb{R}
dx^{\mu_1}\wedge dx^{\mu_2}\wedge dx^{\mu_3}\oplus \bigoplus_{0\leq \mu\leq 3}\Bbb{R} dx^\mu\wedge da
\oplus \Bbb{R}d\pi$.
\item $\displaystyle P^4_{(q,p)}T^*{\cal M}^{Max} = \Bbb{R}\omega\oplus \bigoplus_{0\leq \mu_1<\mu_2\leq 3}\Bbb{R}
dx^{\mu_1}\wedge dx^{\mu_2}\wedge da\oplus \bigoplus_{0\leq \mu\leq 3}\Bbb{R}dx^\mu\wedge d\pi\oplus
\bigoplus_{0\leq \mu\leq 3} \Bbb{R}{\partial \over \partial x^\mu}\iN \theta$.
\end{itemize}
It is worth stressing out the fact that we did not include the
differential of the coordinates $a_\mu$ of $a$ in
$P^1_{(q,p)}T^*{\cal M}^{Max}$. There are strong physical reasons
for that since the gauge potential is not observable. But another
reason is that if we had included the $da_\mu$'s in
$P^1_{(q,p)}T^*{\cal M}^{Max}$, we would not have a copolarization
since $da_\mu\wedge d\pi$ does not satisfy the condition $\forall
X,\widetilde{X}\in {\cal O}_m$, $[X]=[\widetilde{X}]\Rightarrow
b(X)=b(\widetilde{X})$ required. This confirms the {\em agreement}
of the definition of copolarization with physical purposes.

\subsubsection{Results on the dynamics}
We wish here to generalize Proposition \ref{3.2.2.propdyn} to
observable $(p-1)$-forms for $1\leq p<n$. This result actually
justifies the relevance of Definitions \ref{6.1.def4},
\ref{6.1.def5} and \ref{6.1.def6}. Throughout this section  we
assume that $({\cal M},\Omega)$ is equipped with a copolarization.
We start with some technical results. If ${\cal H}$ is a
Hamiltonian function, we recall that we denote by $[X]^{\cal H}$
the class modulo $\sim$ of decomposable $n$-vector fields $X$ such
that $X\iN \Omega =(-1)^nd{\cal H}$.

\begin{lemm}\label{6.2.lem1}
Let $X$ and $\widetilde{X}$ be two decomposable $n$-vectors in $D^n_m{\cal M}$.
If $X \sim \widetilde{X}$ then $\forall 1\leq p\leq n$,
$\forall a\in P^pT^{\star}_m{\cal M}$, 
\begin{equation}\label{6.4.1}
X\Ni a \sim \widetilde{X}\Ni a.
\end{equation}
Hence we can define $[X]\Ni a := [X\Ni a]\in P^{n-p}T{\cal M}$.
\end{lemm}
{\em Proof} ---
This result amounts to the property that for all $0\leq p\leq n$,
$\forall a\in P^pT^{\star}_m{\cal M}$, $\forall b\in P^{n-p}T^{\star}_m{\cal M}$,
\[
\langle X\Ni a, b\rangle = \langle \widetilde{X}\Ni a, b\rangle
\quad \Longleftrightarrow \quad a\wedge b(X) = a\wedge b(\widetilde{X}),
\]
which is true because of $[X] = [\widetilde{X}]$ and $a\wedge b\in P^nT^{\star}_m{\cal M}$. \bbox
\noindent
As a consequence of Lemma \ref{6.2.lem1}, we have the following definition.
\begin{defi}\label{6.2.def1}
Let $F\in \mathfrak{P}^{p-1}{\cal M}$ and ${\cal H}$ a Hamiltonian function.
The pseudobracket $\{{\cal H}, F\}$ is the section of $P^{n-p}T{\cal M}$ defined by
$$\{{\cal H}, F\} := (-1)^{(n-p)p}[X]^{\cal H}\Ni dF.$$
In case $p=n$, $\{{\cal H}, F\}$ is just the scalar function
$[X]^{\cal H}\iN dF = \langle [X]^{\cal H},dF\rangle$.
\end{defi}

\noindent We now prove the basic result relating this notion to the dynamics.

\begin{theo}\label{6.2.thm1}
Let $({\cal M},\Omega)$ be a multisymplectic manifold.
Assume that $1\leq p\leq n$, $1\leq q\leq n$ and $n\leq p+q$.
Let $F\in \mathfrak{P}^{p-1}{\cal M}$ and $G\in \mathfrak{P}^{q-1}{\cal M}$.
Let $\Sigma$ be a slice of codimension $2n-p-q$ and $\Gamma$ a Hamiltonian $n$-curve.
Then for any $(p+q-n)$-vector $Y$ tangent to $\Sigma\cap \Gamma$, we have
\begin{equation}\label{6.2.4}
\{{\cal H}, F\}\iN dG(Y) = (-1)^{(n-p)(n-q)}\{{\cal H}, G\}\iN dF(Y),
\end{equation}
which is equivalent to
\[
\{{\cal H}, F\}\iN dG_{|\Gamma} = (-1)^{(n-p)(n-q)}
\{{\cal H}, G\}\iN dF_{|\Gamma}.
\]
\end{theo}
{\em Proof} --- Proving (\ref{6.2.4}) is equivalent to proving 
\begin{equation}\label{6.2.5}
\langle \{{\cal H},F\}\wedge Y,dG\rangle = (-1)^{(n-p)(n-q)}
\langle \{{\cal H},G\}\wedge Y,dF\rangle.
\end{equation}
We thus need to compute first $\{{\cal H},F\}\wedge Y$.
For that purpose, we use Definition \ref{6.2.def1}:
$\{{\cal H}, F\} = (-1)^{(n-p)(n-q)}[X]^{\cal H}\Ni dF$. Of course it will be more suitable
to use the representant of $[X]^{\cal H}$ which is tangent to $\Gamma$: we let
$(X_1,\cdots ,X_n)$ to be a basis of $T_m\Gamma$ such that
$$X_1\wedge \cdots \wedge X_n =: X \in [X]^{\cal H}$$
Then we can write
$$Y = \sum_{\nu_1<\cdots <\nu_{p+q-n}}T^{\nu_1\cdots \nu_{p+q-n}}
X_{\nu_1}\wedge \cdots \wedge X_{\nu_{p+q-n}}$$

\noindent Now
$$\begin{array}{ccl}
\{{\cal H}, F\} & = & (-1)^{(n-p)p}[X]^{\cal H}\Ni dF\\
 & = & \displaystyle (-1)^{(n-p)p}\sum_{\tiny \begin{array}{c}\mu_1<\cdots <\mu_p\\
\mu_{p+1}<\cdots <\mu_n\end{array}}
\delta^{\mu_1\cdots \mu_n}_{1\cdots n}dF(X_{\mu_1},\cdots ,X_{\mu_p})
X_{\mu_{p+1}}\wedge \cdots \wedge X_{\mu_n},
\end{array}$$
so that\\

\noindent $\displaystyle \{{\cal H}, F\}\wedge Y = (-1)^{(n-p)(p+q-n)}Y\wedge \{{\cal H}, F\}$\\

\noindent $\displaystyle =(-1)^{(n-p)(n-q)}
\sum_{\nu_1<\cdots <\nu_{p+q-n}}
\sum_{\tiny \begin{array}{c}
\mu_1<\cdots <\mu_p\\
\mu_{p+1}<\cdots <\mu_n\end{array}}
T^{\nu_1\cdots \nu_{p+q-n}}
\delta^{\mu_1\cdots \mu_n}_{1\cdots n}$

$$dF(X_{\mu_1},\cdots ,X_{\mu_p})
X_{\nu_1}\wedge \cdots \wedge X_{\nu_{p+q-n}} \wedge
X_{\mu_{p+1}}\wedge \cdots \wedge X_{\mu_n}.
$$
Now $X_{\nu_1}\wedge \cdots \wedge X_{\nu_{p+q-n}} \wedge
X_{\mu_{p+1}}\wedge \cdots \wedge X_{\mu_n}\neq 0$ if and only if it is possible
to complete the family $\{X_{\nu_1},\cdots ,X_{\nu_{p+q-n}}\}$ by
$\{X_{\lambda_1},\cdots ,X_{\lambda_{n-q}}\}$ in such a way that
$\{X_{\nu_1},\cdots ,X_{\nu_{p+q-n}},X_{\lambda_1},\cdots ,X_{\lambda_{n-q}}\}
= \{X_{\mu_1},\cdots ,X_{\mu_p}\}$ and
$\delta^{\nu_1\cdots \nu_{p+q-n}\lambda_1\cdots \lambda_{n-q}}_{\mu_1\cdots \mu_p}\neq 0$.
Hence\\

\noindent $\displaystyle \{{\cal H}, F\}\wedge Y$\\

\noindent $\displaystyle =(-1)^{(n-p)(n-q)}
\sum_{\tiny \begin{array}{c}
\mu_1<\cdots <\mu_p\\
\mu_{p+1}<\cdots <\mu_n\end{array}}
\sum_{\tiny \begin{array}{c}\nu_1<\cdots <\nu_{p+q-n}\\
\lambda_1<\cdots <\lambda_{n-q}\\
\end{array}}
\delta^{\nu_1\cdots \nu_{p+q-n}\lambda_1\cdots \lambda_{n-q}}_{\mu_1\cdots \mu_p}
T^{\nu_1\cdots \nu_{p+q-n}}\delta^{\mu_1..\mu_n}_{1\cdots n}$
$$dF(X_{\nu_1},\cdots ,X_{\nu_{p+q-n}},X_{\lambda_1},\cdots ,X_{\lambda_{n-q}})
X_{\nu_1}\wedge \cdots \wedge X_{\nu_{p+q-n}} \wedge
X_{\mu_{p+1}}\wedge \cdots \wedge X_{\mu_n}$$
\noindent $\displaystyle = (-1)^{(n-p)(n-q)}
\sum_{\tiny \begin{array}{c}
\mu_1<\cdots <\mu_p\\
\mu_{p+1}<\cdots <\mu_n\end{array}}
\sum_{\tiny \begin{array}{c}\nu_1<\cdots <\nu_{p+q-n}\\
\lambda_1<\cdots <\lambda_{n-q}\\
\end{array}}
\delta^{\nu_1\cdots \nu_{p+q-n}\lambda_1\cdots \lambda_{n-q}
\mu_{p+1}\cdots \mu_n}_{1\cdots n}T^{\nu_1\cdots \nu_{p+q-n}}$
$$(X_{\nu_1}\wedge \cdots \wedge X_{\nu_{p+q-n}}\iN dF)(X_{\lambda_1},\cdots ,X_{\lambda_{n-q}})
X_{\nu_1}\wedge \cdots \wedge X_{\nu_{p+q-n}} \wedge
X_{\mu_{p+1}}\wedge \cdots \wedge X_{\mu_n}$$
\noindent $\displaystyle = (-1)^{(n-p)(n-q)}
(-1)^{(n-q)(p+q-n)}X\Ni (Y\iN dF)$\\

\noindent $\displaystyle = (-1)^{(n-q)q}X\Ni (Y\iN dF).$\\

\noindent
We conclude that
$$\begin{array}{ccl}
\langle \{{\cal H}, F\}\wedge Y, dG\rangle & = &
(-1)^{(n-q)q}\langle X,(Y\iN dF)\wedge dG\rangle \\
 & = & \langle X,dG \wedge (Y\iN dF)\rangle \\
 & = & \langle X\Ni dG,Y\iN dF\rangle \\
 & = & (-1)^{(n-q)q}\langle \{{\cal H},G\},Y\iN dF\rangle \\
 & = & (-1)^{(n-q)(n-p)}\langle \{{\cal H},G\}\wedge Y,dF\rangle .
\end{array}$$
So the result follows. \bbox

\begin{coro}\label{6.2.cor1}
Assume the same hypothesis as in Theorem \ref{6.2.thm1}, then we have the following
relations (by decreasing the generality)
\begin{enumerate}
\item If $F\in \mathfrak{P}^{p-1}{\cal M}$ and
$G\in \mathfrak{P}^{n-1}{\cal M}$, then
$$\{{\cal H}, F\}\iN dG_{|\Gamma} = \{{\cal H}, G\}dF_{|\Gamma}$$
\item If $F\in \mathfrak{P}^{p-1}{\cal M}$ and if $G\in
\mathfrak{P}^{n-1}{\cal M}$ is such that $\{{\cal H},G\} = 1$,
then denoting by $\omega := dG$ (a ``volume form'')
$$\{{\cal H}, F\}\iN \omega_{|\Gamma} = dF_{|\Gamma}.$$
\item If $F, G\in \mathfrak{P}^{n-1}{\cal M}$, we recover proposition 1.
\end{enumerate}
\end{coro}
{\em Proof} --- It is a straightforward application of Theorem
\ref{6.2.thm1}. \bbox \noindent {\bf Example 5} --- {\em Consider
a variational problem on maps $u:{\cal X}\longrightarrow {\cal Y}$
as in Example 2, Section 2.2.1.} --- Take $F=y^i$ (a 0-form) and
$G=x^1dx^2\wedge \cdots \wedge dx^n$, in such a way that
$dG=\omega$, the volume form. Then we are in case (ii) of the
corollary: we can compute that $\{{\cal H}, y^i\}\iN \omega =
\sum_\mu
\partial {\cal H}/\partial p^\mu_idx^\mu$ and $\{{\cal H},G\}dy^i=dy^i$.
Hence this implies the relation $dy^i_{|\Gamma}=\sum_\mu
\partial {\cal H}/\partial p^\mu_idx^\mu_{|\Gamma}$.

\section{The ``symmetry'' point of view}
An alternative way to define ``observable forms'' is to suppose
that they are related to infinitesimal symplectomorphisms in a way
analogous to the situation in classical mechanics. This point of
view is more directly related to symmetries and Noether's theorem,
since one can anticipate (correctly) that if the Hamiltonian
function ${\cal H}$ is invariant by an infinitesimal
symplectomorphism then the corresponding observable form is
closed, thus recovering a divergence free vector field. The
advantages of this definition are that we are able to define a
notion of Poisson bracket between such observable forms easily and
that this Poisson bracket is directly related to the one used by
physicists for quantizing fields.

\subsection{Algebraic observable $(n-1)$-forms}

\subsubsection{Definitions}
\begin{defi}\label{3.3.1.def}
Let $m\in {\cal M}$ and $a\in \Lambda^nT^{\star}_m{\cal M}$; $a$
is called {\bf algebraic copolar} if and only if there exists a
unique $\xi\in T_m{\cal M}$ such that $\phi + \xi\iN \Omega =0$.
We denote by $P^n_0T^{\star}_m{\cal M}$ the set of algebraic copolar $n$-forms.\\
\noindent A $(n-1)$-form $F$ on $({\cal M}, \Omega)$ is called
{\bf algebraic observable} $(n-1)$-form if and only if for all
$m\in {\cal M}$, $dF_m\in P^n_0T^{\star}_m{\cal M}$. We denote by
$\mathfrak{P}^{n-1}_0{\cal M}$ the set of all algebraic observable
$(n-1)$-forms.
\end{defi}

\noindent
In other words a $(n-1)$-form $F$ is algebraic observable if and only if
there exists a vector field $\xi$ satisfying $dF+\xi\iN \Omega = 0$. Then
we denote by $\xi_F$ this unique vector field.
A straightforward observation is that any {\em algebraic observable} $(n-1)$-form
satisfies automatically property (\ref{dynaobs}) and so is an {\em observable} $(n-1)$-form.
Actually the following Lemma implies that algebraic observable $(n-1)$-form are
characterized by a property similar to (\ref{dynaobs}) but stronger. (Indeed
${\cal O}^n{\cal M}\subset D^n{\cal M}$ is a submanifold of $\Lambda^nT^{\star}_m{\cal M}$.)

\begin{lemm}\label{3.3.1.lemme}
Let $m\in {\cal M}$ and $\phi\in \Lambda^n_mT^{\star}_m{\cal M}$.
Then $\phi\in P^n_0T^{\star}_m{\cal M}$ if and only if
\begin{equation}\label{4.1.phix}
\forall X,\widetilde{X}\in \Lambda^nT_m{\cal M},\quad
X\iN \Omega = \widetilde{X}\iN \Omega\quad \Longrightarrow \quad \phi(X)=\phi(\widetilde{X}).
\end{equation}
\end{lemm}

\noindent {\em Proof} --- Let us fix some point $m\in {\cal M}$
and let $\phi\in \Lambda^n_mT^{\star}_m{\cal M}$. We consider the
two following linear maps
\[
\begin{array}{ccclccccl}
L: & \Lambda^nT_m{\cal M} & \longrightarrow & T_m^*{\cal M} &
\hbox{ and } & K: & T_m{\cal M} & \longrightarrow &
\left(\Lambda^nT_m{\cal M}\right)^*\\
 & X & \longmapsto & (-1)^nX\iN \Omega & & & \xi & \longmapsto & \xi\iN
 \Omega,
 \end{array}
 \]
 where we used the identification $\left(\Lambda^nT_m{\cal
 M}\right)^*\simeq \Lambda^nT_m^*{\cal M}$ for defining $K$. We
 first observe that $K$ is the adjoint of $L$. Indeed
 \[
 \forall X\in \Lambda^nT_m{\cal M}, \forall \xi\in T_m{\cal
 M},\quad \langle X,K(\xi)\rangle = \xi\iN \Omega(X) = (-1)^nX\iN
\Omega(\xi) = \langle \xi,L(X)\rangle.
\]
Hence since $K$ is one to one (because $\Omega$ is non degenerate)
$L$ is onto. Moreover we can consider the following maps induced
by $L$ and $K$: $[L]: \Lambda^nT_m{\cal M}/\hbox{Ker}L
\longrightarrow T_m^*{\cal M}$ and $[K]: T_m{\cal M}
\longrightarrow \left(\Lambda^nT_m{\cal M}/\hbox{Ker}L\right)^*$.
Again $[K] = [L]^*$ and the fact that $L$ is onto implies that
$[L]$ is a
vector space isomorphism and that $[K]$ is so.\\

\noindent Now observe that the set of $\phi\in
\Lambda^nT^*_m{\cal M}$ which satisfies (\ref{4.1.phix}) coincides
with $\left(\Lambda^nT_m{\cal M}/\hbox{Ker}L\right)^*$. Hence the
conclusion of the Lemma follows from the fact that $[K]: T_m{\cal
M} \longrightarrow \left(\Lambda^nT_m{\cal
M}/\hbox{Ker}L\right)^*$ is an isomorphism. \bbox

\noindent
Hence $\mathfrak{P}^{n-1}_0{\cal M}\subset \mathfrak{P}^{n-1}{\cal M}$.
We wish to single out multisymplectic manifolds where this inclusion is an identity:
\begin{defi}
A multisymplectic manifold $({\cal M}, \Omega)$ is {\bf
pataplectic} if and only if the set of observable $(n-1)$-forms
coincides with the set of algebraic observable $(n-1)$-forms,
i.e.\,$\mathfrak{P}^{n-1}_0{\cal M}= \mathfrak{P}^{n-1}{\cal M}$.
\end{defi}

\noindent We will see in the next paragraph that the
multisymplectic manifold corresponding to the de Donder--Weyl
theory is not pataplectic (if $k\geq 2$). But any open subset of
$\Lambda^nT^*{\cal N}$ is pataplectic, as proved in Section 5
(there we also characterize completely
the set of algebraic observable $(n-1)$-forms).\\

\subsubsection{Example of observable $(n-1)$-forms which are not algebraic observable $(n-1)$-forms}
In order to picture the difference between algebraic and non algebraic observable
$(n-1)$-forms, let us consider the example of the de Donder--Weyl theory
here corresponding to a submanifold of ${\cal M}=\Lambda^nT^{\star}(\Bbb{R}^n\times \Bbb{R}^k)$
(for $n,k\geq 2$) defined in Example 2. We use the same notations as in Example 2.
It is easy to see that the set $\mathfrak{P}^{n-1}_0{\cal M}^{dDW}$
of algebraic observable $(n-1)$-forms coincides with the set of $(n-1)$-forms $F$ on
${\cal M}^{dDW}$ such that, at each point $m\in {\cal M}^{dDW}$, $dF_m$ has the form
\[
dF_m = \left( a^\mu{\partial \over \partial x^\mu} + b^i{\partial \over \partial y^i}\right)
\iN \Omega + f\omega + f_i^\mu\omega^i_\mu
\]
(where we assume summation over all repeated indices).
Now we observe that, by the Pl\"ucker relations,
\[
\forall 1\leq p\leq n,\ \forall X\in D^n{\cal M}^{dDW},\quad
\left(\omega (X)\right)^{p-1}\omega^{i_1\cdots i_p}_{\mu_1\cdots \mu_p}(X) =
\det \left( \omega^{i_\beta}_{\mu_\alpha}(X)\right) _{1\leq \alpha,\beta\leq p},
\]
so it turns out that, if $X\in D^n{\cal M}^{dDW}$ is such that $\omega(X)\neq 0$, then
all the values $\omega^{i_1\cdots i_p}_{\mu_1\cdots \mu_p}(X)$ can be computed from
$\omega(X)$ and $\left( \omega^i_\mu(X)\right) _{1\leq \mu\leq n;1\leq i\leq k}$.\\

\noindent
Hence we deduce that the set of (non algebraic) observable $(n-1)$-forms on
${\cal M}^{dDW}$ contains the set of $(n-1)$-forms $F$ on ${\cal M}^{dDW}$ such that, at each point
$m\in {\cal M}^{dDW}$, $dF_m$ has the form
\[
dF_m = \left( a^\mu{\partial \over \partial x^\mu} + b^i{\partial \over \partial y^i}\right)
\iN \Omega + \sum_{j=1}^n\sum_{i_1<\cdots <i_j}\sum_{\mu_1<\cdots <\mu_j}
f_{i_1\cdots i_p}^{\mu_1\cdots \mu_p}\omega^{i_1\cdots i_p}_{\mu_1\cdots \mu_p}.
\]
Let us denote by $\mathfrak{P}^{n-1}_0\Lambda^nT^*({\cal X}\times {\cal Y})_{|{\cal M}^{dDW}}$
this set. An equivalent definition could be that
$\mathfrak{P}^{n-1}_0\Lambda^nT^*({\cal X}\times {\cal Y})_{|{\cal M}^{dDW}}$
is the set of the restrictions of algebraic observable forms
$\widetilde{F}\in \mathfrak{P}^{n-1}_0\Lambda^nT^*({\cal X}\times {\cal Y})$ on
${\cal M}^{dDW}$ (and this is the reason for this notation). Hence
$\mathfrak{P}^{n-1}{\cal M}^{dDW}\supset
\mathfrak{P}^{n-1}_0\Lambda^nT^*({\cal X}\times {\cal Y})_{|{\cal M}^{dDW}}$. We
will see in Section 5 that the reverse inclusion holds, so that actually
$\mathfrak{P}^{n-1}{\cal M}^{dDW} =
\mathfrak{P}^{n-1}_0\Lambda^nT^*({\cal X}\times {\cal Y})_{|{\cal M}^{dDW}}$, with
${\cal O}_m{\cal M}^{dDW} = \{ X\in D^n_m{\cal M}^{dDW}/\omega(X)\neq 0\}$.

\subsubsection{Invariance properties along pseudo-fibers}

In the following if $\zeta$ is a smooth vector field, we denote by
$e^{s\zeta}$ (for $s\in I$, where $I$ is an interval of $\Bbb{R}$)
its flow mapping. And if $E$ is any subset of ${\cal M}$, we
denote by $E_s:=e^{s\zeta}(E)$ its image by $e^{s\zeta}$.

\begin{lemm}\label{3.3.4.lem}
Let $\Gamma\in {\cal E}^{\cal H}$ be a Hamiltonian $n$-curve and
$\zeta$ be a vector field which is a smooth section of $L^{\cal
H}$ (see Definition \ref{3.3.4.gpd}). Suppose that, for all $s\in
I$, $\Gamma_s$ is a Hamiltonian $n$-curve\footnote{Observe that
this hypothesis is true if ${\cal H}$ is pataplectic invariant,
see Definition \ref{3.3.4.def}}. Let $\Sigma$ be a smooth
$(n-1)$-dimensional submanifold of $\Gamma$ and $F\in
\mathfrak{P}^{n-1}_0{\cal M}$.
If one of the two following hypotheses is satisfied: either\\
(a) $\partial \Sigma =\emptyset$, or\\
(b) $\zeta\iN F= 0$ everywhere, then
\begin{equation}\label{3.3.4.inv}
\forall s\in I,\quad
\int_{\Sigma}F = \int_{\Sigma_s}F.
\end{equation}
i.e.\,the integral of $F$ on the image of $\Sigma$ by $e^{s\zeta}$
does not depend on $s$.
\end{lemm}
\begin{figure}[h]
\begin{center}
\includegraphics[scale=0.5]{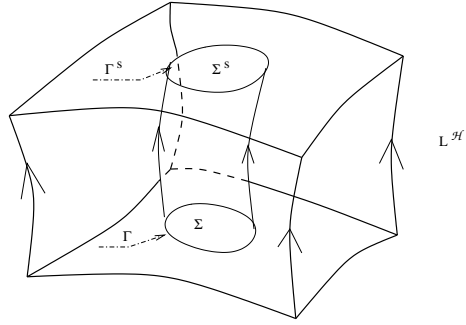}
\caption{\footnotesize Invariance of an observable functional
along the generalized pseudofiber directions as in Lemma
\ref{3.3.4.lem}}
\end{center}
\end{figure}
{\em Proof} --- Let us introduce some extra notations:
$\sigma:I\times \Gamma\longrightarrow {\cal M}$ is the map
$(s,m)\longmapsto \sigma(s,m):=e^{s\zeta}(m)$. Moreover for all
$m\in \Sigma_s\cup \partial \Sigma_s$ we consider a basis
$(X_1,\cdots ,X_n)$ of $T_m\Gamma_s$ such that $X:=X_1\wedge
\cdots \wedge X_n\in [X]^{\cal H}_m$, $(X_2,\cdots ,X_n)$ is a
basis of $T_m\Sigma_s$ and, if $m\in \partial \Sigma_s$,
$(X_3,\cdots ,X_n)$ is a basis of $T_m\partial \Sigma_s$. Lastly
we let $(\theta^1,\cdots ,\theta^n)$ be a basis of
$T^*_m\Gamma_s$, dual of $(X_1,\cdots ,X_n)$. We first note that
\[
\int_{(0,s)\times \partial \Sigma}\sigma^*F = 0,
\]
either because $\partial \Sigma=\emptyset$ (a) or because, if $\partial \Sigma\neq \emptyset$,
this integral is equal to
\[
\int_{ (0,s)\times \partial \Sigma} F_{\sigma(s,m)}
(\zeta,X_3,\cdots ,X_n)ds\wedge \theta^3\wedge \cdots \wedge
\theta^n
\]
which vanishes by (b). Thus
\[
\begin{array}{ccl}
\displaystyle \int_{\Sigma_s}F - \int_{\Sigma}F  =
\int_{\Sigma}\left(e^{s\zeta}\right)^*F-F & = & \displaystyle
\int_{\Sigma}\left(e^{s\zeta}\right)^*F-F
- \int_{(0,s)\times \partial \Sigma}\sigma^*F\\
\displaystyle =\
\int_{\partial \left( (0,s)\times \Sigma\right)} \sigma^*F & = &
\displaystyle \int_{ (0,s)\times \Sigma} d\left(\sigma^*F \right)\\
\displaystyle =\ \int_{ (0,s)\times \Sigma} \sigma^*dF & = &
\displaystyle \int_{ \sigma((0,s)\times \Sigma)} dF_{\sigma(s,m)}
(\zeta,X_2,\cdots ,X_n)ds\wedge \theta^2\wedge \cdots \wedge
\theta^n.
\end{array}
\]
But since $F\in \mathfrak{P}^{n-1}_0{\cal M}$, we have that
\[
\begin{array}{ccc}
dF_{\sigma(s,m)}(\zeta,X_2,\cdots ,X_n) & = & - \Omega(\xi_F,\zeta,X_2,\cdots ,X_n)\\
& = & \Omega(\zeta,\xi_F,X_2,\cdots ,X_n)\\
& =& \langle \xi_F\wedge X_2\wedge \cdots \wedge X_n,\zeta\iN
\Omega \rangle.
\end{array}
\]
Now the key observation is that $\xi_F\wedge X_2\wedge \cdots
\wedge X_n\in T_XD^n_m{\cal M}$ and so the hypothesis of the Lemma
implies that $\langle \xi_F\wedge X_2\wedge \cdots \wedge
X_n,\zeta\iN \Omega \rangle=0$. Hence (\ref{3.3.4.inv}) is
satisfied. \bbox

\noindent {\bf Example 6} --- {\em Algebraic observable
$(n-1)$-forms satisfying the assumption (b) in Lemma
\ref{3.3.4.lem}} --- If ${\cal M} = \Lambda^nT^*{\cal N}$ and if
${\cal H}$ is a Legendre image Hamiltonian then any $(n-1)$-form
$F$ of the type $F=\xi\iN \theta$, where $\xi$ is a vector field
on ${\cal N}$, is algebraic observable (see
\cite{HeleinKouneiher}, \cite{ForgerPauflerRomer}). But as pointed
out in Section 2.2 (definition 2.4) since $L^{\cal H}_m\subset\hbox{Ker}d\Pi_m$ any
vector field $\zeta$ which is a section of $L^{\cal H}$ is
necessarily ``vertical'' and satisfies $\zeta\iN \theta = 0$.
Hence $\zeta\iN F = 0$. Such $(n-1)$-forms $\xi\iN \theta$
correspond to components of the momentum and the energy-momentum
of the field (see \cite{HeleinKouneiher}).

\subsubsection{Poisson brackets between observable $(n-1)$-forms}
There is a natural way to construct a Poisson bracket
$\{\cdot ,\cdot \}:\mathfrak{P}^{n-1}_0{\cal M}$ $\times$
$\mathfrak{P}^{n-1}_0{\cal M}$
$\longmapsto$ $\mathfrak{P}^{n-1}_0{\cal M}$. To each algebraic observable forms
$F,G \in \mathfrak{P}^{n-1}_0{\cal M}$ we associate first the vector fields
$\xi_F$ and $\xi_G$ such that $\xi_F\iN \Omega + dF = \xi_G\iN \Omega + dG = 0$ and then
the $(n-1)$-form
$$\{F,G\} := \xi_F\wedge \xi_G\iN \Omega.$$
It can be shown (see \cite{HeleinKouneiher}) that
$\{F,G\}\in \mathfrak{P}^{n-1}_0{\cal M}$ and that
\[
d\{F,G\} + [\xi_F,\xi_G]\iN \Omega = 0,
\]
where $[\cdot ,\cdot ]$ is the Lie bracket on vector fields.
Moreover this bracket satisfies the Jacobi condition modulo an
exact term\footnote{Note that in case where the multisymplectic
manifold $({\cal M},\Omega)$ is {\bf exact} in the sense of M.
Forger, C. Paufler and H. R\"omer \cite{ForgerPauflerRomer},
i.e.\,if there exists an $n$-form $\theta$ such that $\Omega =
d\theta$ (beware that our sign conventions differ with
\cite{ForgerPauflerRomer}), an alternative Poisson bracket can be
defined:
\[
\{F,G\}_\theta:= \{F,G\} + d(\xi_G\iN F - \xi_F\iN G + \xi_F\wedge \xi_G\iN \theta).
\]
Then this bracket satisfies the Jacobi identity (in particular with a right hand side
equal to 0), see \cite{ForgerRomer}, \cite{ForgerPauflerRomer}.} (see \cite{HeleinKouneiher})
\[
\{\{F,G\},H\} + \{\{G,H\},F\} + \{\{H,F\},G\} =
d(\xi_F\wedge \xi_G\wedge \xi_H\iN \Omega).
\]

\noindent
As an application of this definition, for any slice $\Sigma$ of codimension 1
we can define a Poisson bracket
between the observable functionals $\int_{\Sigma}F$ and $\int_{\Sigma}G$ by
$\forall \Gamma \in {\cal E}^{\cal H}$,
\[
\left\{ \int_{\Sigma}F, \int_{\Sigma}G\right\} (\Gamma) :=
\int_{\Sigma\cap \Gamma}\{F,G\}.
\]
If $\partial \Gamma = \emptyset$, it is clear that this Poisson bracket satisfies the
Jacobi identity. Computations in \cite{Kijowski1}, \cite{Kanatchikov1}, \cite{HeleinKouneiher}
show that this Poisson bracket coincides with the Poisson bracket of
the standard canonical formalism used for quantum field theory.\\

\noindent One can try to extend the bracket between forms in
$\mathfrak{P}^{n-1}_0{\cal M}$ to forms in
$\mathfrak{P}^{n-1}{\cal M}$ through different strategies:
\begin{itemize}
\item By exploiting the relation
\[
\{F,G\} = \xi_F\iN dG = - \xi_G\iN dF,
\]
which holds for all $F,G\in \mathfrak{P}^{n-1}_0{\cal M}$. A natural definition
is to set:
\[
\forall F\in \mathfrak{P}^{n-1}_0{\cal M},\,
\forall G\in \mathfrak{P}^{n-1}{\cal M},\quad
\{F,G\} = - \{G,F\} := \xi_F\iN dG.
\]
In \cite{HeleinKouneiher} we call this operation an {\bf external}
Poisson bracket. \item If we know that there is an embedding
$\iota :{\cal M}\longrightarrow \widehat{\cal M}$, into a higher
dimensional pataplectic manifold $(\widehat{\cal
M},\widehat{\Omega})$, and that (i)
$\mathfrak{P}^{n-1}_0\widehat{\cal M}
=\mathfrak{P}^{n-1}\widehat{\cal M}$, (ii) the pull-back mapping
$\mathfrak{P}^{n-1}_0\widehat{\cal M}\longrightarrow
\mathfrak{P}^{n-1}{\cal M}: \widehat{F}\longmapsto
\iota^*\widehat{F}$ is --- modulo the set of closed $(n-1)$-forms
on $\widehat{\cal M}$ which vanish on ${\cal M}$ --- an
isomorphism. Then there exists a unique Poisson bracket on
$\mathfrak{P}^{n-1}{\cal M}$ which is the image of the Poisson
bracket on
$\mathfrak{P}^{n-1}_0\widehat{\cal M}$.\\
This situation is achieved for instance if ${\cal M}$ is a submanifold
of $\Lambda^nT^*{\cal N}$, a situation which arises after a Legendre transform.
This will lead basically to the same structure as the external Poisson bracket. In more
general cases the question of extending ${\cal M}$ into $\widehat{\cal M}$ is relatively
subtle and is discussed in the paper \cite{HeleinKouneiher1.2}.
\end{itemize}

\subsection{Algebraic observable $(p-1)$-forms}
It is worth to ask about the relevant definition of algebraic
observable $(p-1)$-forms. Indeed a first possibility is to
generalize directly Definition \ref{3.3.1.def}: we would say that
the $(p-1)$-form $F$ is algebraic observable if there exists a
$(1+n-p)$-multivector field $\xi_F$ such that $dF + \xi_F\iN
\Omega = 0$. Note that in this case $\xi_F$ is not unique in
general. This approach has been proposed by I. Kanatchikov in
\cite{Kanatchikov1} and \cite{Kanatchikov2}. It allows us to
define a nice notion of Poisson bracket between such forms and
leads to a structure of graded Lie algebra. However this
definition has not good dynamical properties and in particular
there is no analogue of Theorem \ref{6.2.thm1} for such forms. In
other words our definition \ref{6.1.def5} of (non algebraic)
observable $(p-1)$-forms results from the search for the more
general hypothesis for Theorem \ref{6.2.thm1} to be true, but
$(p-1)$-forms which are observable according to Kanatchikov are
not observable in the sense of our Definition \ref{6.1.def5}. Here
we prefer to privilege the dynamical properties instead trying to
generalize Noether's theorem to $(p-1)$-forms.

\subsubsection{Definitions and basic properties}

We simply adapt Definitions \ref{6.1.def4}, \ref{6.1.def5} and
\ref{6.1.def6} by replacing $P^nT^*_m{\cal M}$ of Definition
\ref{3.2.1.def} by its subset $P^n_0T^*_m{\cal M}$ of Definition
\ref{3.3.1.def}: it leads to the notions of {\bf algebraic
copolarization} $P_0^*T^*{\cal M}$, of the set
$\mathfrak{P}_0^{p-1}{\cal M}$ of {\bf algebraic observable
$(p-1)$-forms} and of {\bf algebraic polarization} $P_0^*T{\cal
M}$. Of course in the case of a pataplectic manifold algebraic and
non algebraic notions coincide.\\

\noindent A consequence of these definitions is that for any
$1\leq p\leq n$, for any algebraic observable $(p-1)$-forms $F\in
\mathfrak{P}_0^{p-1}{\cal M}$ and for any $\phi \in
P^{n-p}_0T^{\star}_m{\cal M}$, there exists a unique vector
$\xi_F(\phi)\in T_m{\cal M}$ such that $\phi\wedge dF +
\xi_F(\phi)\iN \Omega = 0$. We thus obtain a linear mapping
\begin{equation}\label{6.1.xi}
\begin{array}{cccc}
\xi_F: & P^{n-p}_0T^{\star}_m{\cal M} & \longrightarrow & T_m{\cal M}\\
 & \phi & \longmapsto & \xi_F(\phi).
\end{array}
\end{equation}
Hence we can associate to $F$ the tensor field $\xi_F$. By duality
between $P^{n-p}_0T^{\star}_m{\cal M}$ and $P^{n-p}_0T_m{\cal M}$,
$\xi_F$ can also be identified with a section of the bundle
$P^{n-p}_0T{\cal M}\otimes _{\cal M}T{\cal M}$.\\

\noindent
Moreover an alternative definition of the pseudobracket (see Definition \ref{6.2.def1})
can be given using the tensor field $\xi_F$ defined by (\ref{6.1.xi}).
\begin{lemm}\label{6.2.lem2}
For any Hamiltonian function ${\cal H}$ and any $F\in \mathfrak{P}_0^{p-1}{\cal M}$,
we have
\begin{equation}\label{6.2.2}
\{{\cal H}, F\} = - \xi_F\iN d{\cal H},
\end{equation}
where the right hand side is the section of $P^{n-p}_0T{\cal M}$ defined by
\begin{equation}\label{6.2.3}
\langle \xi_F\iN d{\cal H}, \phi\rangle :=
\xi_F(\phi)\iN d{\cal H},\quad \forall \phi \in P^{n-p}_0T^{\star}{\cal M}.
\end{equation}
\end{lemm}
{\em Proof} --- Starting from Definition \ref{6.2.def1},
we have $\forall \phi \in P^{n-p}_0T^{\star}{\cal M}$,
\[
\begin{array}{ccl}
\langle \{{\cal H},F\}, \phi\rangle & = &
(-1)^{(n-p)p}\langle [X]^{\cal H}\Ni dF, \phi\rangle \\
 & = & (-1)^{(n-p)p}\langle [X]^{\cal H}, dF\wedge \phi\rangle \\
 & = & \langle [X]^{\cal H}, \phi\wedge dF\rangle  \\
 & = & - \langle [X]^{\cal H}, \xi_F(\phi)\iN \Omega\rangle \\
 & = & - (-1)^n\xi_F(\phi)\iN [X]^{\cal H}\iN \Omega\\
 & = & - \xi_F(\phi)\iN d{\cal H}.
\end{array}
\]
\bbox

\noindent The price we have to pay is that the notion of Poisson
bracket between $(p-1)$-forms is now a much more delicate task
than in the framework of Kanatchikov. This question is the subject
of the next paragraph.

\subsubsection{Brackets between algebraic observable
$(p-1)$-forms}

We now consider algebraic observable $(p-1)$-forms for $1\leq
p\leq n$ and discuss the possibility of defining a Poisson bracket
between these observable forms, which could be relevant for
quantization. This is slightly more delicate than for forms of
degree $n-1$ and the definitions proposed here are based on
empirical observations. We first assume a further hypothesis on
the copolarization (which is
satisfied on $\Lambda^nT^*{\cal N}$).\\

\noindent We first recall a definition which was given in
\cite{HK1a}: given any $X=X_1\wedge \cdots \wedge X_n\in
D^n_m{\cal M}$ and any form $a\in T^*_m{\cal M}$ we will write
that $a_{|X}\neq 0$  if and only if $(a(X_1),\cdots ,a(X_n))\neq
0$. We will say that a function $f\in {\cal C}^1({\cal M},\R)$ is
1-regular if and only if we have
\begin{equation}\label{4.1/2.differe}
\forall m\in {\cal M}, \forall X\in [X]^{\cal H}_m,\quad df_{m|X}
\neq 0.
\end{equation}
\noindent {\bf Hypothesis on $\mathfrak{P}^1_0{\cal M}$} --- {\em
We suppose that any 1-regular function $f\in {\cal C}^1({\cal
M},\R)$ satisfies\footnote{these assumptions are actually
satisfied in $\mathfrak{P}^{n-1}{\cal M}$}
\begin{enumerate}
\item $f\in \mathfrak{P}^1_0{\cal M}$

\item For all infinitesimal symplectomorphism  $\xi\in
\mathfrak{sp}_0{\cal M}$, $\xi\iN df = 0$.
\end{enumerate}}

\noindent Now let $1\leq p,q\leq n$ and $F\in
\mathfrak{P}^{p-1}_0{\cal M}$ and $G\in \mathfrak{P}^{q-1}_0{\cal
M}$ and let us analyze what condition should satisfy the bracket
$\{F,G\}$. We will consider smooth functions $f^1,\cdots
,f^{n-p}$, $g^1,\cdots ,g^{n-q}$ and $t$ on ${\cal M}$. We assume
that all these functions are 1-regular and that $df^1_m\wedge
\cdots \wedge df^{n-p}_m\wedge dg^1_m\wedge \cdots \wedge
dg^{n-q}_m\neq 0$. Then, because of hypothesis (i),
\[
\widetilde{F}:= df^1\wedge \cdots \wedge df^{n-p}\wedge F,\quad
 \widetilde{G}:= dg^1\wedge \cdots \wedge
dg^{n-q}\wedge G\in\mathfrak{P}^{n-1}_0{\cal M}.
\]
Lastly let $\Gamma$ be a Hamiltonian $n$-curve and $\Sigma$ be a
level set of $t$. Then
\begin{equation}\label{6.4.bracket.n-1n-1}
\left\{ \int_{\Sigma}\widetilde{F},\int_{\Sigma}\widetilde{G}
\right\}(\Gamma) = \left( \int_{\Sigma}\left\{
\widetilde{F},\widetilde{G}\right\} \right)(\Gamma) =
\int_{\Sigma\cap \Gamma}\left\{
\widetilde{F},\widetilde{G}\right\}.
\end{equation}
We now suppose that the functions $f:=(f^1,\cdots ,f^{n-p})$ and
$g:=(g^1,\cdots ,g^{n-q})$ concentrate around submanifolds denoted
respectively by $\widehat{\gamma}_f$ and $\widehat{\gamma}_g$ of
codimension $n-p$ and $n-q$ respectively. More precisely we
suppose that $df^1\wedge \cdots \wedge df^{n-p}$ (resp.
$dg^1\wedge \cdots \wedge dg^{n-q}$) is zero outside a tubular
neighborhood of $\widehat{\gamma}_f$ (resp. of
$\widehat{\gamma}_g$) of width $\varepsilon$ and that the integral
of $df^1\wedge \cdots \wedge df^{n-p}$ (resp. $dg^1\wedge \cdots
\wedge dg^{n-q}$) on a disc submanifold of dimension $n-p$ (resp.
$n-q$) which cuts transversally $\widehat{\gamma}_f$ (resp.
$\widehat{\gamma}_g$) is equal to 1. Moreover we suppose that
$\widehat{\gamma}_f$ and $\widehat{\gamma}_g$ cut transversally
$\Sigma\cap \Gamma$ along submanifolds denoted by $\gamma_f$ and
$\gamma_g$ respectively. Then, as $\varepsilon\rightarrow 0$, we
have
\[
\int_{\Sigma\cap \Gamma}df^1\wedge \cdots \wedge df^{n-p}\wedge F
\rightarrow \int_{\Sigma\cap \widehat{\gamma}_f\cap \Gamma}F,
\quad \int_{\Sigma\cap \Gamma}dg^1\wedge \cdots \wedge
dg^{n-q}\wedge G \rightarrow \int_{\Sigma\cap
\widehat{\gamma}_g\cap \Gamma}G.
\]
This tells us that the left hand side of
(\ref{6.4.bracket.n-1n-1}) is an approximation for
\[
\left\{ \int_{\Sigma\cap \widehat{\gamma}_f}F, \int_{\Sigma\cap
\widehat{\gamma}_g}G \right\}(\Gamma).
\]
\begin{figure}[h]
\begin{center}
\includegraphics[scale=0.5]{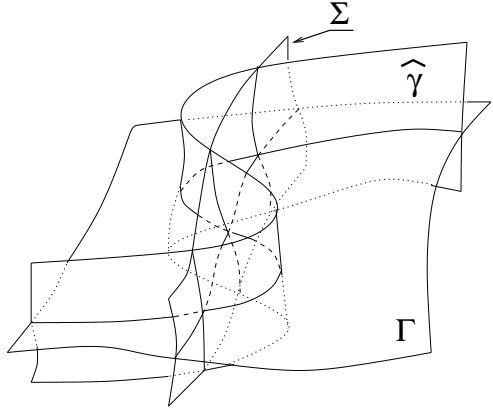}
\caption{\footnotesize Intersection of $\Gamma$,
$\widehat{\gamma}$ and $\Sigma$}
\end{center}
\end{figure}
We now want to compute what is the limit of the right hand side of
(\ref{6.4.bracket.n-1n-1}). Using the hypothesis (ii) we have
\[
\begin{array}{ccl}
\{\widetilde{F},\widetilde{G}\} & = & \xi_{\widetilde{F}}\iN
d\left(
dg^1\wedge \cdots \wedge dg^{n-q}\wedge G\right) \\
& = & (-1)^{n-q}\xi_{\widetilde{F}}\iN dg^1\wedge \cdots \wedge dg^{n-q}\wedge dG\\
& = & dg^1\wedge \cdots \wedge dg^{n-q}\wedge \left(
\xi_{\widetilde{F}}\iN dG\right).
\end{array}
\]
And we have similarly:
\[
\{\widetilde{F},\widetilde{G}\} = - df^1\wedge \cdots \wedge
df^{n-p}\wedge \left( \xi_{\widetilde{G}}\iN dF\right).
\]
We now use the following result.
\begin{lemm}\label{6.4.lemmetechnique}
Let $\phi\in \Lambda^{n-1}T^*_m{\cal M}$ with $\phi\neq 0$ and
$1\leq p,q\leq n$ such that $p+q\geq n+1$. Suppose that there
exists $2n-p-q$ linearly independent 1-forms $a^1,\cdots
,a^{n-p},b^1,\cdots ,b^{n-q}\in T^*_m{\cal M}$, $\alpha\in
\Lambda^{p-1}T^*_m{\cal M}$ and $\beta\in \Lambda^{q-1}T^*_m{\cal
M}$ such that $\phi = a^1\wedge \cdots \wedge a^{n-p}\wedge
\alpha$ and $\phi = b^1\wedge \cdots \wedge b^{n-q}\wedge \beta$.
Then there exists $\chi \in \Lambda^{p+q-n-1}T^*_m{\cal M}$ such
that $\phi = a^1\wedge \cdots \wedge a^{n-p}\wedge b^1\wedge
\cdots \wedge b^{n-q}\wedge \chi$. This $\chi$ is not unique in
general and is defined modulo forms in the ideal in
$\Lambda^*T^*_m{\cal M}$ spanned by the $a_j$'s and the $b_j$'s.
However it is a unique real scalar if $p+q = n+1$.
\end{lemm}
{\em Proof} --- This is a consequence of Proposition 1.4 in
\cite{bcggg}. The idea is based on the observation that
$a^1,\cdots ,a^{n-p},b^1,\cdots ,b^{n-q}$ are in $\{a\in
T^*_m{\cal M}/a\wedge \phi=0\}$.\bbox \noindent We deduce from
Lemma \ref{6.4.lemmetechnique} that there exist a form $\chi \in
\Lambda^{p+q-n-1}T^*_m{\cal M}$ (not unique a priori) such that
$\{\widetilde{F},\widetilde{G}\} = df^1\wedge \cdots \wedge
df^{n-p}\wedge dg^1\wedge \cdots \wedge dg^{n-q}\wedge \chi$. We
thus require that
\begin{equation}\label{6.4.implicite}
\{\widetilde{F},\widetilde{G}\} = df^1\wedge \cdots \wedge
df^{n-p}\wedge dg^1\wedge \cdots \wedge dg^{n-q}\wedge \{F,G\}.
\end{equation}
This does not characterize completely $\{F,G\}$, unless $p+q =
n+1$, the case $\{F,G\}$ where is a scalar. We can now write the
right hand side of (\ref{6.4.bracket.n-1n-1}) as
\[
\int_{\Sigma\cap \Gamma} df^1\wedge \cdots \wedge df^{n-p}\wedge
dg^1\wedge \cdots \wedge dg^{n-q}\wedge \{F,G\}.
\]
Letting $\varepsilon \rightarrow 0$, and assuming that
$\widehat{\gamma}_f$ and $\widehat{\gamma}_g$ cross transversally
this integral converges to
\[
\int_{\Sigma\cap \widehat{\gamma}_f\cap \widehat{\gamma}_g\cap
\Gamma} \{F,G\},
\]
so that we have
\[
\left\{ \int_{\Sigma\cap \widehat{\gamma}_f}F, \int_{\Sigma\cap
\widehat{\gamma}_g}G \right\}(\Gamma) = \int_{\Sigma\cap
\widehat{\gamma}_f\cap \widehat{\gamma}_g\cap \Gamma} \{F,G\}.
\]
Here the intersection $\Sigma\cap \widehat{\gamma}_f\cap
\widehat{\gamma}_g\cap \Gamma$ is oriented by assuming that $X\Ni
dt \wedge df\wedge dg$ is oriented positively, if $X\in [X]^{\cal
H}$ orients positively $T_m\Gamma$. Hence if we had started with
$\int_{\Sigma\cap \Gamma}\{\widetilde{F},\widetilde{G}\} = -
\int_{\Sigma\cap \Gamma}\{\widetilde{G},\widetilde{F}\}$ we would
have obtained $-\int_{\Sigma\cap \widehat{\gamma}_g\cap
\widehat{\gamma}_f\cap \Gamma} \{G,F\} = -(-1)^{(n-p)(n-q)}
\int_{\Sigma\cap \widehat{\gamma}_f\cap \widehat{\gamma}_g\cap
\Gamma}\{G,F\}$. Since the resulting brackets should coincide we
deduce that
\[
\{F,G\} + (-1)^{(n-p)(n-q)}\{G,F\} = 0.
\]

\noindent {\bf Conclusion} --- In two cases we can guess a more
direct definition of $\{F,G\}$. First when one of the two forms
$F$ or $G$ is in $\mathfrak{P}^{n-1}_0{\cal M}$, let us say $F\in
\mathfrak{P}^{p-1}_0{\cal M}$ and $G\in \mathfrak{P}_0^{n-1}{\cal
M}$: then we let
\begin{equation}\label{5.5.1.p-1n-41}
\{F,G\} := -\xi_G\iN dF.
\end{equation}
This is the idea of external bracket as in Paragraph 4.1.4. We
remark that if $f^1,\cdots ,f^{n-p}$ are in $\mathfrak{P}^1_0{\cal
M}$ and are such that $\xi_G\iN df^1\wedge \cdots \wedge df^{n-p}
= 0$, then $df^1\wedge \cdots \wedge df^{n-p}\wedge F\in
\mathfrak{P}^{n-1}_0{\cal M}$ and
\[
\{df^1\wedge \cdots \wedge df^{n-p}\wedge F, G\} = df^1\wedge
\cdots \wedge df^{n-p}\wedge \{F,G\}
\]
so that the requirement (\ref{6.4.implicite}) is satisfied.\\

\noindent Second if $F\in \mathfrak{P}^{p-1}_0{\cal M}$, $G\in
\mathfrak{P}^{q-1}_0{\cal M}$, where  $1<p,q<n$ and $p+q=n+1$:
then $\{F,G\}$ is just a scalar and so is characterized
by (\ref{6.4.implicite}).\\

\noindent {\bf Example 7} --- {\em Sigma models} --- Let ${\cal
M}:= \Lambda^nT^{\star}({\cal X}\times {\cal Y})$ as in Section 2.
For simplicity we restrict ourself to the de Donder--Weyl
submanifold ${\cal M}^{dDW}$ (see Section 2.3), so that the
Poincar\'e-Cartan form is $\theta = e\omega + p^\mu_idy^i\wedge
\omega_\mu$ and the multisymplectic form is $\Omega = d\theta$.
Let $\phi$ be a function on ${\cal X}$ and consider the observable
0-form $y^i$ (for $1\leq i\leq k$) and the observable $(n-1)$-form
$P_{j,\phi}:= \phi(x)\partial / \partial y^j\iN \theta.$ Then
$\xi_{P_{j,\phi}} = \phi\partial / \partial y^j -
p^\mu_j\left(\partial \phi/ \partial x^\mu\right)
\partial / \partial e$ and thus $\{P_{j,\phi},y^i\}= \xi_{P_{j,\phi}}\iN dy^i
= \delta^i_j \phi$. It gives the following bracket for observable
functionals
\[
\left\{ \int_{\Sigma}P_{j,\phi},\int_{\Sigma\cap
\widehat{\gamma}}y^i\right\}(\Gamma) = \int_{\Sigma\cap
\widehat{\gamma}\cap \Gamma}\delta^i_j \phi(x) = \delta^i_j
\sum_{m\in \Sigma\cap \widehat{\gamma}\cap \Gamma}
\hbox{sign}(m)\phi(m),
\]
where $\Sigma$, $\widehat{\gamma}$ and $\Gamma$ are supposed to
cross transversally
and $\hbox{sign}(m)$ accounts for the orientation of their intersection points.\\

\noindent {\bf Example 4''} --- {\em Maxwell equations} --- In
this case we find that, for all functions $f,g_1,g_2:{\cal
M}^{Max}\longrightarrow \Bbb{R}$ whose differentials are proper on
${\cal O}_m$, $\{df\wedge \pi,dg_1\wedge dg_2\wedge a\} = df\wedge
dg_1\wedge dg_2$. We hence deduce that $\{\pi,a\} =1$: these forms
are canonically conjugate. We deduce the following bracket for
observable functionals
\[
\left\{ \int_{\Sigma\cap \widehat{\gamma}_f}\pi,\int_{\Sigma\cap
\widehat{\gamma}_g}a\right\}(\Gamma) =  \sum_{m\in \Sigma\cap
\widehat{\gamma}_f\cap \widehat{\gamma}_g\cap \Gamma}
\hbox{sign}(m),
\]
where $\Sigma\cap \widehat{\gamma}_f\cap \Gamma$ is a surface and
$\Sigma\cap \widehat{\gamma}_g\cap \Gamma$ is a curve in the
three-dimensional space $\Sigma\cap \Gamma$. Note that this
conclusion was achieved by I.\,Kanatchikov with its definition of
bracket $\{\pi,a\}_{Kana}:=\xi_\pi\iN da$, where $\xi_\pi \in
\Lambda^2T^*_m{\cal M}^{Max}$ is such that $\xi_\pi\iN \Omega =
d\pi$. But there by choosing $\xi_\pi=(1/2)\sum_\mu (\partial
/\partial a_\mu)\wedge (\partial /\partial x^\mu)$ one finds (in
our convention) that $\{\pi,a\}_{Kana} = n/2$ ($= 2$, if $n=4$).
So the two brackets differ (the new bracket in this paper differs
also from the one that we proposed in \cite{HeleinKouneiher}).

\section{The case of $\Lambda^nT^{\star}{\cal N}$}
We will here study algebraic and non algebraic observable
$(n-1)$-forms in $\Lambda^nT^{\star}{\cal N}$ and prove in
particular that $\Lambda^nT^{\star}{\cal N}$ is pataplectic. This
part can be seen as a complement of an analysis of other
properties of $\Lambda^nT^{\star}{\cal N}$ given in \cite{HK1a}.
In particular we use the same notations: since we are interested
here in local properties of ${\cal M}$, we will use local
coordinates $m=(q,p)=(q^\alpha,p_{\alpha_1\cdots \alpha_n})$ on
${\cal M}$, and the multisymplectic form reads
$\Omega=\sum_{\alpha_1<\cdots <\alpha_n}dp_{\alpha_1\cdots
\alpha_n}\wedge dq^{\alpha_1}\wedge \cdots  \wedge dq^{\alpha_n}$.
For $m=(q,p)$, we write
\[
d_q{\cal H}:= \sum_{1\leq \alpha\leq n+k}{\partial {\cal H}\over \partial q^\alpha}dq^\alpha,\quad
d_p{\cal H}:= \sum_{1\leq \alpha_1<\cdots <\alpha_n\leq n+k}
{\partial {\cal H}\over \partial p_{\alpha_1\cdots \alpha_n}}dp_{\alpha_1\cdots \alpha_n},
\]
so that $d{\cal H} = d_q{\cal H} + d_p{\cal H}$.

\subsection{Algebraic and non algebraic observable $(n-1)$-forms coincide}
We show here that $\left(\Lambda^nT^{\star}{\cal N},\Omega\right)$
is a pataplectic manifold.
\begin{theo}\label{4.3.theopata}
If ${\cal M}$ is an open subset of $\Lambda^nT^{\star}{\cal N}$, then
$\mathfrak{P}^{n-1}_0{\cal M}=\mathfrak{P}^{n-1}{\cal M}$.
\end{theo}
{\em Proof} --- We already know that $\mathfrak{P}^{n-1}_0{\cal M}\subset \mathfrak{P}^{n-1}{\cal M}$.
Hence we need to prove the reverse inclusion. So in the following we consider some $m\in {\cal M}$ and
a form $a\in P_m^n{\cal M}$ and we will prove that there exists a vector
field $\xi$ on ${\cal M}$ such that $a=\xi\iN \Omega$. We write ${\cal O}_m{\cal M}:=
{\cal O}^a_m{\cal M}$.\\

\noindent
{\em Step 1} --- We show that given $m=(q,p)\in {\cal M}$ it is
possible to find $n+k$ vectors $(\widetilde{Q}_1,\cdots ,\widetilde{Q}_{n+k})$ of $T_m{\cal M}$ such that,
if $\Pi_*(\widetilde{Q}_\alpha) =: Q_\alpha$ (the image of $\widetilde{Q}_\alpha$ by the
map $\Pi:{\cal M}\longrightarrow {\cal N}$), then $(Q_1,\cdots ,Q_{n+k})$ is
a basis of $T_q{\cal N}$ and $\forall (\alpha_1,\cdots , \alpha_n)$
s.t.\,$1\leq \alpha_1<\cdots < \alpha_n\leq n+k$,
$\widetilde{Q}_{\alpha_1}\wedge \cdots \wedge \widetilde{Q}_{\alpha_n}\in {\cal O}_m{\cal M}$.\\

\noindent
This can be done by induction by using the fact that ${\cal O}_m{\cal M}$ is dense
in $D^n_m{\cal M}$. We start from any family of vectors $(\widetilde{Q}^0_1,\cdots ,\widetilde{Q}^0_{n+k})$
of $T_m{\cal M}$ such that $(Q^0_1,\cdots ,Q^0_{n+k})$ is a basis of $T_q{\cal N}$
(where $Q^0_\alpha:= \Pi_*(\widetilde{Q}^0_\alpha)$). We then order the ${(n+k)!\over n!k!}$
multi-indices $(\alpha_1,\cdots ,\alpha_n)$ such that $1\leq \alpha_1<\cdots < \alpha_n\leq n+k$
(using for instance the dictionary rule).
Using the density of ${\cal O}_m{\cal M}$ we can perturb slightly
$(\widetilde{Q}^0_1,\cdots ,\widetilde{Q}^0_{n+k})$ into
$(\widetilde{Q}^1_1,\cdots ,\widetilde{Q}^1_{n+k})$ in such a way that for instance
$\widetilde{Q}^1_1\wedge \cdots \wedge \widetilde{Q}^1_n\in {\cal O}_m{\cal M}$
(assuming that $(1,\cdots ,n)$ is the smallest index). Then we perturb further
$(\widetilde{Q}^1_1, \cdots ,\widetilde{Q}^1_{n+k})$ into
$(\widetilde{Q}^2_1, \cdots , \widetilde{Q}^2_{n+k})$ in such a way
that $\widetilde{Q}^2_1\wedge \cdots \wedge \widetilde{Q}^2_{n-1}\wedge
\widetilde{Q}^2_{n+1}\in {\cal O}_m{\cal M}$ (assuming that $(1,\cdots ,n-1,n+1)$ is the next
one). Using the fact that ${\cal O}_m{\cal M}$ is open we can do it in such a way
that we still have $\widetilde{Q}^2_1\wedge \cdots \wedge \widetilde{Q}^2_n\in {\cal O}_m{\cal M}$.
We proceed further until the conclusion is reached.\\

\noindent
In the following we choose local coordinates around $m$ in such a way that
$\widetilde{Q}_\alpha = \partial _\alpha + \sum_{1\leq \alpha_1<\cdots < \alpha_n\leq n+k}
P_{\alpha,\alpha_1\cdots \alpha_n}\partial ^{\alpha_1\cdots \alpha_n}$.\\

\noindent
{\em Step 2} --- We choose a multi-index $(\alpha_1,\cdots ,\alpha_n)$
with $1\leq \alpha_1<\cdots < \alpha_n\leq n+k$ and define the
set ${\cal O}_m^{\alpha_1\cdots \alpha_n}{\cal M}:={\cal O}_m{\cal M}
\cap D_m^{\alpha_1\cdots \alpha_n}{\cal M}$, where
\[
D_m^{\alpha_1\cdots \alpha_n}{\cal M}:=
\left\{ X_1\wedge \cdots \wedge X_n\in D^n_m{\cal M}/
\forall \mu, X_\mu = {\partial \over \partial q^{\alpha_\mu}}
+ \sum_{1\leq \beta_1<\cdots < \beta_n\leq n+k}
X_{\mu,\beta_1\cdots \beta_n}{\partial \over \partial p^{\beta_1\cdots \beta_n}}\right\}.
\]
We want to understand the consequences of the relation
\begin{equation}\label{4.3.property}
\forall X,\widetilde{X}\in {\cal O}_m^{\alpha_1\cdots \alpha_n}{\cal M},\quad
X\iN \Omega = \widetilde{X}\iN \Omega\quad \Longrightarrow \quad a(X) = a(\widetilde{X}).
\end{equation}
Note that ${\cal O}_m^{\alpha_1\cdots \alpha_n}{\cal M}$ is open and non empty (since by the previous
step, $\widetilde{Q}_{\alpha_1}\wedge \cdots \wedge \widetilde{Q}_{\alpha_n}\in
{\cal O}_m^{\alpha_1\cdots \alpha_n}{\cal M}$).
We also observe that, on $D_m^{\alpha_1\cdots \alpha_n}{\cal M}$, $X\longmapsto X\iN \Omega$
and $X\longmapsto a(X)$ are respectively an affine function and a polynomial
function of the coordinates variables $X_{\mu,\beta_1\cdots \beta_n}$. Thus the
following result implies that actually ${\cal O}_m^{\alpha_1\cdots \alpha_n}{\cal M} =
D_m^{\alpha_1\cdots \alpha_n}{\cal M}$.
\begin{lemm}\label{4.3.technique1}
Let $N\in \Bbb{N}$ and let $P$ be a polynomial on $\Bbb{R}^N$ and
$f_1,\cdots, f_p$ be affine functions on $\Bbb{R}^N$. Assume that
there exists some $x_0\in \Bbb{R}^N$ and a neighborhood $V_0$ of
$x_0$ in $\Bbb{R}^N$ such that
\[
\forall x,\widetilde{x}\in V_0,\quad \hbox{if }\forall j = 1,\cdots ,p,
\ f_j(x) = f_j(\widetilde{x}), \quad \hbox{then }P(x) = P(\widetilde{x}).
\]
Then this property is true on $\Bbb{R}^N$.
\end{lemm}
{\em Proof} --- We can assume without loss of generality that the functions
$f_j$ are linear and also choose coordinates on $\Bbb{R}^N$ such that
$f_j(x) = x^j$, $\forall j = 1,\cdots ,p$. Then the assumption means that,
on $V_0$, $P$ does not depend on $x^{p+1},\cdots x^N$. Since $P$ is a polynomial
we deduce that $P$ is a polynomial on the variables $x^1,\cdots ,x^p$ and so the
property is true everywhere. \bbox
\noindent
{\em Step 3} --- Without loss of generality we will also assume in the following
that $(\alpha_1,\cdots ,\alpha_n) = (1,\cdots, n)$ for simplicity. We
shall denote by $m^I$ all coordinates $q^\alpha$ and $p_{\alpha_1\cdots \alpha_n}$,
so that we can write
\[
a = \sum_{I_1<\cdots <I_n}A_{I_1\cdots I_n}dm^{I_1}\wedge \cdots \wedge dm^{I_n}.
\]
We will prove that if $(I_1,\cdots ,I_n)$ is a multi-index such that
\begin{itemize}
\item $\{m^{I_1},\cdots ,m^{I_n}\}$ contains at least two distinct coordinates
of the type $p_{\alpha_1\cdots \alpha_n}$ and
\item $\{m^{I_1},\cdots ,m^{I_n}\}$ does not contain any $q^\alpha$, for
$n+1\leq\alpha \leq n+k$
\end{itemize}
then $A_{I_1\cdots I_n} = 0$. Without loss of generality we can suppose that $\exists p\in \Bbb{N}$
such that $1\leq p\leq n-2$ and
\[
m^{I_1} = q^1,\cdots ,m^{I_p} = q^p\quad \hbox{and}\quad
m^{I_{p+1}},\cdots ,m^{I_n}\in \{p_{\alpha_1\cdots \alpha_n}/1\leq\alpha_1<\cdots <\alpha_n\leq n+k\}.
\]
We test property (\ref{4.3.property}) specialized to the case where $X=X_1\wedge \cdots \wedge X_n$ with
\[
X_\mu = {\partial \over \partial q^\mu} + \sum_{j=p+1}^nX_\mu^{I_j}
{\partial \over \partial m^{I_j}},\quad \forall \mu = 1,\cdots ,n.
\]
Then
\begin{equation}\label{4.3.determinant}
a(X) = A_{I_1\cdots I_n}\left|\begin{array}{cccccc}
1 & \cdots & 0 & 0 & \cdots & 0\\
\vdots & \ddots & \vdots & \vdots & & \vdots \\
0 & \cdots & 1 & 0 & \cdots & 0\\
X_1^{I_{p+1}} & \cdots & X_p^{I_{p+1}} & X_{p+1}^{I_{p+1}} & \cdots & X_n^{I_{p+1}}\\
\vdots & & \vdots & \vdots &  & \vdots \\
X_1^{I_n} & \cdots & X_p^{I_n} & X_{p+1}^{I_n} & \cdots & X_n^{I_n}
\end{array}\right|
 = A_{I_1\cdots I_n}\left|\begin{array}{ccc}
X_{p+1}^{I_{p+1}} & \cdots & X_n^{I_{p+1}}\\
\vdots &  & \vdots \\
X_{p+1}^{I_n} & \cdots & X_n^{I_n}
\end{array}\right| .
\end{equation}
Note that we can write any $X\in D_m^{1\cdots n}{\cal M}$ as $X=X_1\wedge \cdots \wedge X_n$
with
\[
X_\mu = {\partial \over \partial q^\mu} + E_\mu{\partial \over \partial p_{1\cdots n}}
+ \sum_{\beta = n+1}^{n+k}M_{\mu,\beta} + R_\mu,
\]
where $M_{\mu,\beta}:= \sum_{\nu = 1}^n(-1)^{n+\nu}M^\nu_{\mu,\beta}
\partial ^{1\cdots \widehat{\nu}\cdots n\beta}$ and
$R_\mu:= \sum_{(\alpha_1,\cdots ,\alpha_n)\in I^{**}}X_{\mu,\alpha_1\cdots \alpha_n}
\partial ^{\alpha_1\cdots \alpha_n}$. And then
\[
(-1)^nX\iN \Omega = dp_{1\cdots n} - \sum_{\mu=1}^nE_\mu dq^\mu -
\sum_{\beta=n+1}^{n+k}\left(\sum_{\mu=1}^nM_{\mu,\beta}^\mu\right) dq^\beta.
\]
Within our specialization this leads to the following {\em key
observation}\footnote{Remark that each of the $n-p$ last lines in
the $n\times n$ determinant in (\ref{4.3.determinant}) is either
$(E_1,\cdots ,E_n)$ or of the type $(M_{1,\beta}^\nu,\cdots
,M_{m,\beta}^\nu)$ or $(X_{1,\alpha_1\cdots \alpha_n},\cdots
,X_{n,\alpha_1\cdots \alpha_n})$, for $(\alpha_1,\cdots
,\alpha_n)\in I^{**}$.}: {\bf at most one} line $(X_1^{I_j},\cdots
,X_n^{I_j})$ ( for $p+1\leq j\leq n$) in the $n\times n$
determinant in (\ref{4.3.determinant}) is a function of $X\iN
\Omega$ (for $m^{I_j}=p_{1\cdots n}$). In all other lines number
$\nu$, where $p+1\leq \nu\leq n$ and $\nu\neq j$, there is {\bf at
most} one component $X^{I_\nu}_\mu$ which is a function of $X\iN
\Omega$. All the other components are independent of $X\iN
\Omega$. Thus we have the following alternative.
\begin{enumerate}
\item $\{m^{I_{p+1}},\cdots ,m^{I_n}\}$ does not contain $p_{1\cdots n}$
(i.e.\,the line $(E_1,\cdots ,E_n)$ does not appear in the $n\times n$ determinant in (\ref{4.3.determinant})),
or
\item $\{m^{I_{p+1}},\cdots ,m^{I_n}\}$ contains $p_{1\cdots n}$
(i.e.\,one of the lines is $(E_1,\cdots ,E_n)$)
\end{enumerate}
{\em Case} (i) --- Then the right hand side determinant in
(\ref{4.3.property}) is a polynomial of degree $n-p\geq 2$. Thus
we can find a monomial in this determinant of the form
$X^{I_{p+1}}_{\sigma(p+2)}\cdots X^{I_n}_{\sigma(n)}$ (where
$\sigma$ is a substitution of $\{p+1,\cdots ,n\}$) where each
variable is independent of $X\iN \Omega$.
Hence in order to achieve (\ref{4.3.property}) we must have $A_{I_1\cdots I_n} = 0$.\\
{\em Case} (ii) --- We assume w.l.g.\,that $m^{I_{p+1}} =
p_{1\cdots n}$. We freeze the variables $X^{I_{p+1}}_\mu$
(i.e.\,$E_\mu$) suitably and specialize again property
(\ref{4.3.property}) by letting free only the variables
$X^{I_j}_\mu$ for $p+2\leq j\leq n$ and $1\leq \mu \leq n$. Two
subcases occur: if $p < n-2$ then we choose $X^{I_{p+1}}_\mu =
\delta^{p+1}_\mu$. Then we are reduced to a situation quite
similar to the first case and we can conclude using the same
argument (this time with a determinant which is a monomial of degree $n-1-p\geq 2$).\\
If $p= n-2$ then $a(X) = A_{I_1\cdots I_n}\left(
X_n^{I_n} X_{n-1}^{I_{n-1}} - X_{n-1}^{I_n} X_n^{I_{n-1}}\right)$. If the knowledge of
$X\iN \Omega$ prescribes $X_n^{I_n}$ then by the key observation $X_{n-1}^{I_n}$ is free and by
choosing $X^{I_{n-1}}_\mu = \delta^n_\mu$ we obtain $a(X) = - A_{I_1\cdots I_n}X_{n-1}^{I_n}$.
If $X\iN \Omega$ prescribes $X_{n-1}^{I_n}$ then $X_n^{I_n}$ is free and by
choosing $X^{I_{n-1}}_\mu = \delta^{n-1}_\mu$ we obtain $a(X) = A_{I_1\cdots I_n}X_n^{I_n}$.
In both cases we must have $A_{I_1\cdots I_n} = 0$ in order to have (\ref{4.3.property}).\\

\noindent
{\em Conclusion} --- Steps 2 and 3 show that, on ${\cal O}_m^{1\cdots n}{\cal M}$,
$X\longmapsto a(X)$ is an affine function on the variables $X_{\mu,\beta_1\cdots \beta_n}$. Then
by standard results in linear algebra (\ref{4.3.property}) implies that,
$\forall X\in {\cal O}_m^{1\cdots n}{\cal M}$,
$a(X)$ is an affine combination of the components of $X\iN \Omega$. By repeating this
step on each ${\cal O}_m^{\alpha_1\cdots \alpha_n}{\cal M}$ we deduce the conclusion.
\bbox
\begin{theo}\label{4.3.theodDW}
Assume that ${\cal N} = {\cal X}\times {\cal Y}$, ${\cal M} = \Lambda^nT^*{\cal N}$ and consider
${\cal M}^{dDW}$ to be the submanifold of $\Lambda^nT^*{\cal N}$ as defined
in Paragraph 2.3.1 equipped with the multisymplectic form $\Omega^{dDW}$
which is the restriction of $\Omega$ to ${\cal M}^{dDW}$. Then $\mathfrak{P}^{n-1}{\cal M}^{dDW}$
coincides with $\mathfrak{P}^{n-1}_0\Lambda^nT^*({\cal X}\times {\cal Y})_{|{\cal M}^{dDW}}$,
the set of the restrictions of algebraic observable $(n-1)$-forms of
$({\cal M},\Omega)$ to ${\cal M}^{dDW}$.
\end{theo}
{\em Proof} --- The fact that $\mathfrak{P}^n{\cal M}^{dDW}$
contains all the restrictions of algebraic observable
$(n-1)$-forms of $({\cal M},\Omega)$ to ${\cal M}^{dDW}$ was
observed in Paragraph 4.2.1. The proof of the reverse inclusion
follows the same strategy as the proof of Theorem
\ref{4.3.theopata} and is left to the reader. \bbox

\subsection{All algebraic observable $(n-1)$-forms}
We conclude this section by giving the expression of all algebraic
observable $(n-1)$-forms on an open subset of
$\Lambda^nT^{\star}{\cal N}$. For details see \cite{HK1a}. First
the general expression of an infinitesimal symplectomorphism is
$\Xi = \chi + \overline{\xi}$, where
\begin{equation}\label{4.3.chixi}
\chi := \sum_{\beta_1<\cdots <\beta_n}\chi_{\beta_1\cdots \beta_n}(q)
{\partial \over \partial p_{\beta_1\cdots \beta_n}} \quad \hbox{ and }\quad
\overline{\xi}:= \sum_\alpha \xi^\alpha(q){\partial \over \partial q^\alpha}
- \sum_{\alpha,\beta}{\partial \xi^\alpha\over \partial q^\beta}(q)\Pi^\beta_\alpha,
\end{equation}
and where the coefficients $\chi_{\beta_1\cdots \beta_n}$ are so that $d(\chi\iN \Omega)= 0$,
$\xi:= \sum_\alpha \xi^\alpha(q){\partial \over \partial q^\alpha}$
is an arbitrary vector field on ${\cal N}$,
and $\Pi^\beta_\alpha:= \sum_{\beta_1<\cdots <\beta_n}\sum_\mu \delta^\beta_{\beta_\mu}
p_{\beta_1\cdots \beta_{\mu-1}\alpha\beta_{\mu+1}\cdots \beta_n}
{\partial \over \partial p_{\beta_1\cdots \beta_n}}$.\\

\noindent
As a consequence any algebraic observable $(n-1)$-form $F$ can be written as
$F=Q^\zeta + P_\xi$, where
\[
Q^\zeta = \sum_{\beta_1<\cdots <\beta_{n-1}}
\zeta_{\beta_1\cdots \beta_{n-1}}(q)dq^{\beta_1}\wedge \cdots  \wedge dq^{\beta_{n-1}}
\quad \hbox{and}\quad P_\xi = \xi\iN \theta.
\]
Then $\chi\iN \Omega = -dQ^\zeta$ and
$\overline{\xi}\iN \Omega = - dP_\xi$.

\noindent Lastly we let\footnote{recall that $\mathfrak{sp}_0{\cal
M}$ is the set of all symplectomorphisms of $({\cal M},\Omega)$
(see Definition \ref{2.1.def3})} $\mathfrak{sp}_Q{\cal M}$ to be
the set of infinitesimal symplectomorphisms of the form $\chi$
(with $\chi\iN \Omega$ closed) and $\mathfrak{sp}_P{\cal M}$ those
of the form $\overline{\xi}$ (for all vector fields $\xi\in
\Gamma({\cal M},T{\cal M})$) as defined in (\ref{4.3.chixi}). Then
one can observe that $\mathfrak{sp}_0{\cal M}
=\mathfrak{sp}_P{\cal M}\ltimes \mathfrak{sp}_Q{\cal M}$ (see
\cite{HeleinKouneiher1}).

\section{Dynamical observable forms and functionals}
One question is left: to make sense of the Poisson bracket of two
observable functionals supported on different slices. This is
essential in an Einstein picture (classical analogue of the
Heisenberg picture) which seems unavoidable in a completely
covariant theory. One possible answer rests on the notion of {\em
dynamical} observable forms (in contrast with kinematic observable
functionals). To be more precise let $\Sigma$ and $\Sigma'$ be two
different slices of codimension 1 and $F$ and $G$ be two algebraic
observable $(n-1)$-forms and
let us try to define the Poisson bracket between $\int_{\Sigma}F$ and $\int_{\Sigma'}G$.\\

\noindent
One way is to express one of the two observable functionals, say $\int_{\Sigma'}G$, as
an integral over $\Sigma$. This can be achieved for all slices $\Sigma$ and $\Sigma'$ which are
cobordism equivalent, i.e.\,such that there exists a smooth
domain ${\cal D}$ in ${\cal M}$ with $\partial {\cal D} = \Sigma' - \Sigma$, and if
$dG_{|\Gamma} = 0$, $\forall \Gamma \in {\cal E}^{\cal H}$.
Then indeed
\begin{equation}\label{3.1.cobord}
\int_{\Sigma\cap \Gamma}G - \int_{\Sigma'\cap \Gamma}G =
\int_{\partial {\cal D}\cap \Gamma}G
=\int_{{\cal D}\cap \Gamma}dG = 0,
\end{equation}
so that
\[
\int_{\Sigma'}G = \int_{\Sigma}G\quad \hbox{on}\quad {\cal E}^{\cal H}.
\]
Thus we are led to the following.

\begin{defi}
A {\em dynamical observable $(n-1)$-form} is an observable form
$G\in \mathfrak{P}^{n-1}{\cal M}$ such that
$$ \{{\cal H}, G\} = 0$$
\end{defi}
Indeed Corollary \ref{3.2.2.corodyn} implies immediately that if $G$ is a dynamical
observable $(n-1)$-form then $dG_{|\Gamma} = 0$ and hence
(\ref{3.1.cobord}) holds. As a consequence if
$F$ is any observable $(n-1)$-form and $G$ is a dynamical
observable $(n-1)$-form (and if one of both is an algebraic one), then we can state
$$\left\{ \int_{\Sigma}F, \int_{\Sigma'}G\right\} :=
\left\{ \int_{\Sigma}F, \int_{\Sigma}G\right\} .$$
The concept of dynamical observable form is actually more or less the one used by J. Kijowski
in \cite{Kijowski1}, since his theory corresponds to working on the
restriction of $({\cal M}, \Omega)$ on the hypersurface ${\cal H} = 0$.\\

\noindent
Hence we are led to the question of characterizing dynamical observable $(n-1)$-forms.
(We shall consider mostly algebraic observable forms.)
This question was already investigated for some particular case in \cite{Kijowski1}
(and discussed in \cite{GoldschmidtSternberg})
and the answer was a (surprising) deception: as long as the variational problem
is linear (i.e.\,the Lagrangian is a quadratic function of all variables)
there are many observable functionals (basically all smeared integrals of fields using test functions
which satisfy the Euler-Lagrange equation), but as soon as
the problem is non linear the choice of dynamical observable forms is dramatically reduced
and only global dynamical observable exists. For instance for a non nonlinear
scalar field theory with $L(u, du) = {1\over 2}(\partial _tu)^2
-|\nabla u|^2 + {m^2\over 2}u^2 + {\lambda \over 3}u^3$, the only dynamical
observable forms $G$ are those for which $\xi_G$ is a generator of the Poincar\'e
group. One can also note that in general dynamical observable forms correspond
to momentum or energy-momentum observable functionals.\\

\noindent
Several possibilities may be considered to go around this difficulty. If the variational
problem can be seen as a deformation of a linear one (i.e.\,of a free field theory)
then it could be
possible to construct a perturbation theory, leading to Feynman type expansions for classical
fields. For an example of such a theory, see \cite{perturb} and \cite{Dika}.
Another interesting direction would be to
explore completely integrable systems. We present here a third alternative, which relies
on symmetries and we will see on a simple example how the purpose of constructing dynamical observable forms leads
naturally to gauge theories.\\

\noindent {\bf Example 8} --- {\em Complex scalar fields} --- We
consider on the set of maps $\varphi:\Bbb{R}^n\longrightarrow
\Bbb{C}$ the variational problem with Lagrangian
\[
L_0(\varphi,d\varphi) = {1\over 2}\eta^{\mu\nu}
{\partial \overline{\varphi}\over \partial x^\mu}
{\partial \varphi\over \partial x^\nu} + V\left( {|\varphi|^2\over 2}\right) =
{1\over 2}\eta^{\mu\nu}\left(\
{\partial \varphi^1\over \partial x^\mu}{\partial \varphi^1\over \partial x^\nu} +
{\partial \varphi^2\over \partial x^\mu}{\partial \varphi^2\over \partial x^\nu}\right)
+ V\left( {|\varphi|^2\over 2}\right) .
\]
Here $\varphi = \varphi^1+i\varphi^2$. We consider the multisymplectic manifold ${\cal M}_0$, with
coordinates $x^\mu$, $\phi^1$, $\phi^2$, $e$, $p^\mu_1$ and $p^\mu_2$ and the multisymplectic
form $\Omega_0= de\wedge \omega + dp^\mu_a\wedge d\phi^a\wedge \omega_\mu$ (which is the differential
of the Poincar\'e-Cartan form $\theta_0:= e\omega + p^\mu_ad\phi^a\wedge \omega_\mu$).
Then the Hamiltonian is
\[
{\cal H}_0(x,\phi,e,p) = e + {1\over 2}\eta_{\mu\nu}(p^\mu_1p^\nu_1
+p^\mu_2p^\nu_2) - V\left( {|\phi|^2\over 2}\right) .
\]
We look for $(n-1)$-forms $F_0$ on ${\cal M}_0$ such that
\begin{equation}\label{7.2.1}
dF_0 + \xi_{F_0}\iN \Omega_0 = 0,\quad \hbox{for some vector field}\quad
\xi_{F_0},
\end{equation}
\begin{equation}\label{7.2.2}
d{\cal H}_0(\xi_{F_0}) = 0.
\end{equation}
The analysis of this problem can be dealt by looking for all vector fields
\[
\xi_0 = X^\mu(x,\phi,e,p){\partial \over \partial x^\mu} + \Phi^a(x,\phi,e,p)
{\partial \over \partial \phi^a}
+ E(x,\phi,e,p) {\partial \over \partial e} + P^\mu_a(x,\phi,e,p)
{\partial \over \partial p^\mu_a}.
\]
satisfying (\ref{7.2.1}) and (\ref{7.2.2}).
For simplicity we will assume that $X^\mu = 0$ (this will exclude stress-energy tensor observable
forms $X^\mu{\partial \over \partial x^\mu}\iN \theta_0$, for $X^\mu$ constant). Then we find two cases:\\

\noindent
If $V(|\phi|^2/2)$ is quadratic in $\phi$, i.e.\,if $V(|\phi|^2/2) = m^2|\phi|^2/2$,
then Equations (\ref{7.2.1}) and (\ref{7.2.1}) have the solutions
\[
\xi_0 = \lambda \vec{j}_0
+ U^a(x){\partial \over \partial \phi^a}
- \left(p^\mu_a{\partial U^a\over \partial x^\mu}(x) + \delta_{ab}LU^a(x)\phi^b\right){\partial \over \partial e}
+ \eta^{\mu\nu}\delta_{ab}{\partial U^a\over \partial x^\mu}(x){\partial \over \partial p^\mu_b},
\]
where $\lambda$ is a real constant,
\[
\vec{j}_0 := \left( \phi^2{\partial \over \partial \phi^1} - \phi^1{\partial \over \partial \phi^2}\right)
+ \left(p^\mu_2{\partial \over \partial p^\mu_1} - p^\mu_1{\partial \over \partial p^\mu_2}\right),
\]
$L:= - \eta^{\mu\nu}{\partial ^2\over \partial x^{\mu}\partial x^{\nu}}$
and $U^1$ and $U^2$ are arbitrary solutions of the linear equation $LU + m^2U = 0$.
Then $F_0=U^ap^\mu_a\omega_\mu - \eta^{\mu\nu}\left(
{\partial U^1\over \partial x^\nu}\phi^1 + {\partial U^2\over \partial x^\nu}\phi^2
\right) \omega_\mu + \lambda \left( p^\mu_1\phi^2 - p^\mu_2\phi^1\right) \omega_\mu$.\\

\noindent
However if $V'$ is not a constant, then system (\ref{7.2.1}) and (\ref{7.2.1}) has only the solutions
$\xi_0 = \lambda \vec{j}_0$ and the resulting dynamical observable $(n-1)$-form is
$F_0 = \lambda(p^\mu_1\phi^2 - p^\mu_2\phi^1)\omega_\mu$, which
corresponds to the global charge due to the $U(1)$ invariance of the Lagrangian.\\

\noindent
For instance we would like to replace $\lambda$ by a smooth function $\psi$ of $x$,
i.e.\,to look at $F = \psi(x)(p^\mu_1\phi^2 - p^\mu_2\phi^1)\omega_\mu$. These are
non dynamical algebraic observable $(n-1)$-forms since we have
$dF_1 + \widetilde{\xi}\iN \Omega_0 = 0$, where
$\widetilde{\xi}:= \psi\vec{j}_0 - (p^\mu_1\phi^2 - p^\mu_2\phi^1){\partial \psi\over \partial x^\mu}
{\partial \over \partial e}$,
but $d{\cal H}_0(\widetilde{\xi}) = - (p^\mu_1\phi^2 - p^\mu_2\phi^2){\partial \psi\over \partial x^\mu}\neq 0$.\\

\noindent
Now in order to enlarge the set of dynamical observable forms, an idea is to further incorporate the
gauge potential field $A:= A_\mu dx^\mu$ and consider the Lagrangian
\[
L_1(\varphi,A,d\varphi):= {1\over 2}\eta^{\mu\nu}
\left( \overline{ {\partial \varphi\over \partial x^\mu} + iA_\mu \varphi}\right)
\left( {\partial \varphi\over \partial x^\nu} + iA_\nu \varphi\right) - {1\over 4}\eta^{\mu\lambda}\eta^{\nu\sigma}
F_{\mu\nu}F_{\lambda\sigma} + V\left( {|\varphi|^2\over 2}\right),
\]
where $F_{\mu\nu}:={\partial A_\nu\over \partial x^\mu} - {\partial A_\mu\over \partial x^\nu}$.
It is invariant under gauge transformations $\varphi\longmapsto e^{i\theta}\varphi$,
$A\longmapsto A - d\theta$. Note that we did incorporate an energy
for the gauge potential $A$. We now consider the multisymplectic manifold
${\cal M}_1$ with coordinates $x^\mu$, $\phi^1$, $\phi^2$, $e$,
$p^\mu_1$, $p^\mu_2$, $a_\mu$ and $p^{\mu\nu}$.
The multisymplectic form is: $\Omega_1= de\wedge \omega + dp^\mu_a\wedge d\phi^a\wedge \omega_\mu
- (da_\lambda\wedge dx^\lambda )\wedge ({1\over 2}dp^{\mu\nu}\wedge \omega_{\mu\nu})$.
The Hamiltonian is then
\[
{\cal H}_1(x,\phi,a,e,p) = e + {1\over 2}\eta_{\mu\nu}(p^\mu_1p^\nu_1 + p^\mu_2p^\nu_2)
+ (p^\mu_1\phi^2 - p^\mu_2\phi^1)a_\mu
- {1\over 4}\eta_{\mu\lambda}\eta_{\nu\sigma}p^{\mu\nu}p^{\lambda\sigma}
- V\left( {|\phi|^2\over 2}\right) .
\]
The gain is that we may now consider the algebraic observable $(n-1)$-form
\[
F_1:= \psi(x) (p^\mu_1\phi^2 - p^\mu_2\phi^1)\omega_\mu
-{1\over 2}p^{\mu\nu}d\psi \wedge \omega_{\mu\nu}.
\]
where $\psi$ is any smooth function of $x$. We indeed still have on the one hand
$dF_1 = - \xi_1 \iN \Omega_1$,
where
\[
\xi_1 := \psi\vec{j}_0 - (p^\mu_1\phi^2 - p^\mu_2\phi^1){\partial \psi\over \partial x^\mu}
{\partial \over \partial e}
+ {\partial \psi\over \partial x^\mu}{\partial \over \partial a_\mu}.
\]
Then $d{\cal H}_1(\xi_{F_1}) = 0$. Thus $F_1$ is a dynamical
observable $(n-1)$-form.

\end{document}